\documentclass{JHEP3}
\usepackage{amsmath,amssymb,epsfig}

\newcommand{\be}{\begin{equation}}
\newcommand{\ee}{\end{equation}}
\newcommand{\beq}{\begin{equation}}
\newcommand{\eeq}{\end{equation}}
\newcommand{\ba}{\begin{eqnarray}}
\newcommand{\ea}{\end{eqnarray}}
\newcommand{\bea}{\begin{eqnarray}}
\newcommand{\eea}{\end{eqnarray}}

\newcommand{\reef}[1]{(\ref{#1})}

\newcommand{\ff}{f_{\infty}}

\def\vt#1#2#3 {{\vartheta[{#1 \atop  #2}](#3\vert \tau)}}

\title{Lovelock theories, holography and the fate of the viscosity bound}

\author{Xi\'an O. Camanho\\
\sl Department of Particle Physics and IGFAE, University of Santiago de Compostela, E-15782 Santiago de Compostela, Spain\\\vskip-4mm
\email{xian.otero@rai.usc.es}}

\author{Jos\'e D. Edelstein\\
\sl Department of Particle Physics and IGFAE, University of Santiago de Compostela, E-15782 Santiago de Compostela, Spain\\\vskip-3mm
\sl Centro de Estudios Cient\'\i ficos, Valdivia, Chile\\\vskip-4mm
\email{jose.edelstein@usc.es}}

\author{Miguel F. Paulos\\
\sl Laboratoire de Physique Th\'eorique et Hautes Energies, CNRS UMR 7589,\\ Universit\'e Pierre et Marie Curie, 4 place Jussieu, 75252 Paris Cedex 05, France\\\vskip-4mm
\email{mfp24@cam.ac.uk}}

\bigskip
\bigskip
\abstract{We consider Lovelock theories of gravity in the context of AdS/CFT. We show that, for these theories, causality violation on a black hole background can occur well in the interior of the geometry, thus posing more stringent constraints than were previously found in the literature. Also, we find that instabilities of the geometry can appear for certain parameter values at any point in the geometry, as well in the bulk as close to the horizon. These new sources of causality violation and instability should be related to CFT features that do not depend on the UV behavior. They solve a puzzle found previously concerning unphysical negative values for the shear viscosity that are not ruled out solely by causality restrictions. We find that, contrary to previous expectations, causality violation is not always related to positivity of energy. Furthermore, we compute the bound for the shear viscosity to entropy density ratio of supersymmetric conformal field theories from $d=4$ till $d=10$ -- {\it i.e.}, up to quartic Lovelock theory --, and find that it behaves smoothly as a function of $d$. We propose an approximate formula that nicely fits these values and has a nice asymptotic behavior when $d \to \infty$ for any Lovelock gravity. We discuss in some detail the latter limit. We finally argue that it is possible to obtain increasingly lower values for $\eta/s$ by the inclusion of more Lovelock terms.}


\begin{document}

\section{Introduction}

The AdS/CFT correspondence has provided a tool to study hydrodynamical aspects of quantum field theories at strong coupling. This was particularly timely due to the advent of experiments that prompted the exploration of QCD in a region of phase space where it displays such behavior. One of the most striking predictions of AdS/CFT had to do with the shear viscosity to entropy density ratio, $\eta/s$, of strongly coupled plasmas. Interestingly enough, it was found that there is a universal value for this ratio, $\eta/s = 1/4\pi$, for theories whose gravity dual is governed by the Einstein--Hilbert action \cite{Policastro2001a}, regardless of the matter content, the number of supersymmetries, the existence or not of a conformal symmetry, and even the spacetime dimensionality. On the other hand, all measured values for this ratio in any quantum relativistic system are above this value. This led to speculations that $\eta/s \geq 1/4\pi$ might be an exact statement in quantum relativistic systems, the so-called KSS bound conjecture \cite{Kovtun2005}. Indeed, it is possible to provide a hand waving argument, relying on a quasiparticle description of the plasma, that links the KSS bound to the Heisenberg uncertainty principle. This argument is questionable, though, since there are convincing hints supporting a non-quasiparticle description of strongly interacting plasmas.

The KSS bound conjecture has been thoroughly scrutinized for many years (see, for example, \cite{Dobado2008} for a recent review). It turned out to be the case, however, that when quantum corrections are included in the gravitational action, under the form of curvature squared terms, the $\eta/s$ ratio can be smaller than the KSS bound \cite{Brigante2008,Kats2009}. In particular, there are string theory constructions where this is the case \cite{Kats2009,Buchel2009}. On general grounds, the coefficient of the curvature squared terms must be small. However, for the particular combination given by the Gauss-Bonnet (GB) term, one may consider finite values of the coefficient, as the gravitational theory is then two-derivative. In this case the ratio is modified to $\eta/s = 1/4\pi (1- 4 \lambda)$, $\lambda$ being the appropriately normalized GB coupling, and the KSS bound is violated whenever $\lambda$ is positive \cite{Brigante2008,Kats2009}. Even if {\it a priori} the addition of a GB term would lead to an arbitrary violation of the KSS bound, it turns out that causality constraints arise preventing the possibility of going to arbitrarily low values of $\eta/s$ \cite{Brigante2008a}. In the case of 4d CFTs, for instance, this imposes the constraint $\lambda \leq 9/100$, which reduces the minimum value of $\eta/s$ by a factor of $16/25$.\footnote{It is interesting to mention at this point that the link between causality violation and the viscosity bound does not apply for thermal theories undergoing a low temperature phase transition \cite{Buchel2010b}.}

The study of a possible bound is interesting both from a theoretical standpoint as well as from a phenomenological one -- it has been found experimentally that the quark-gluon plasma created in relativistic heavy ion collisions appears to have a very low shear viscosity to entropy density ratio \cite{Romatschke2007b}. A similar result was also found recently in a radically different context: that of strongly correlated ultracold atomic Fermi gasses in the so-called {\it unitarity} limit \cite{Schafer2009}. Both systems have measured values of $\eta/s$ compatible with $1/4\pi$. On the other hand, from a theoretical standpoint, the causality constraints coming from the behavior of the geometry near the boundary, that rule the attainability of lower values of $\eta/s$, have a beautiful holographic dual in the CFT side: they arise from positivity of energy conditions \cite{Hofman2008,Buchel2009e,Hofman2009}. The perfect matching of these two quite fundamental restrictions on a physically sensible theory constitutes a striking check of the AdS/CFT correspondence. 

Whether this is still valid for higher-dimensional CFTs became a natural question that was subsequently answered in a series of papers \cite{Boer2010a,Camanho2010,Buchel2010a}. Summarizing, it turned out that GB theories lead to a violation of the KSS bound in any spacetime dimensionality, and causality constraints exactly match positivity of energy bounds on the CFT side.  The fact that this matching is valid regardless of the dimensionality is puzzling and seems to provide clues on possible non-stringy versions of AdS/CFT. Our knowledge of higher-dimensional CFTs is, however, too poor yet to push these arguments forward. In higher dimensions one has in general the choice of including other Lovelock terms in the action. These are higher order in curvature that still lead to second order equations of motion \cite{Lovelock1971}. A first step in the analysis of these theories was done in \cite{Boer2010,Camanho2010a}. There it was found that causality and positivity of energy constraints still match perfectly, provided a conjectural relation between field theory parameters and Lovelock couplings was true (something that is proved in the present article). Still, something new happens here. Since there are now more free parameters available, it turns out that causality alone cannot prevent the $\eta/s$ ratio from becoming arbitrarily small (or even negative).

In this paper we analyze in detail the fate of the $\eta/s$ ratio in AdS/CFT restricted to the case of Lovelock theories. We start by describing, in section 2, the lagrangians, AdS vacua and black brane solutions of such theories. The latter are determined by solving a simple polynomial of the same order as the highest Lovelock term in the action. We then establish the holographic dictionary for such theories in section 3, by computing the central charge $C_T$ that controls the stress tensor $2$-point function, and the two independent quantities, $t_2$ and $t_4$, appearing in the $3$-point function \cite{Hofman2008}. With these parameters we can jump into the CFT side and establish positivity of energy bounds tantamount to a series of constraints on the couplings of the gravitational lagrangian. 

We then consider graviton fluctuations about the black hole background for a general Lovelock theory. In the high-momentum limit, one can recast their equations of motion as Schr\"odinger-type equations with surprisingly simple effective potentials, which also have the interpretation of the effective speed of light of the graviton at various radii. Near the boundary, demanding causality of graviton propagation leads to bounds which exactly match those of positivity of energy. Close to the horizon, in turn, we show that unstable modes can appear if the parameters of the theory are not tuned properly. This leads to another set of bounds discussed in section 4, which for the cubic Lovelock theory in seven dimensions prevents the $\eta/s$ ratio from becoming arbitrarily small. In the case of higher order Lovelock theories, a new phenomenon occurs\footnote{This phenomenon already takes place for the cubic Lovelock theory. It leads to a less stringent (thus irrelevant) constraint in $d=7$, but is crucial  to prevent arbitrarily small values of $\eta/s$ in $8 \leq d \leq 10$ theories. When $d>10$, causality at the boundary is enough to discard this possibility in cubic Lovelock gravity \cite{Camanho2010a}.} which is one of the main results of this paper: causality violation and plasma instabilities may arise in the interior of the geometry. These are thoroughly discussed in section 5. They should be related to analogous unhealthy properties arising in the CFT plasma far from the UV, whose identification and full characterization is left as an open problem. They should correspond, however, to very different physical situations: while we may expect that causality violation corresponds to some illness of the boundary theory, instabilities should be related to either plasma instabilities or to issues such as not being expanding around the true vacuum\footnote{We thank Rob Myers for clarifying comments on these points.}. Anyway, interestingly enough, this shows that causality violation is not necessarily a phenomenon arising in the UV and it is not generically dual to positivity of energy in the CFT.

The attentive reader has already realized that higher order Lovelock gravities imply higher-dimensional theories. There is a second more subtle point that should be presently raised. The analysis performed in this article deals with higher order curvature corrections whose coupling constants do not need to be neccesarily small. Both conditions are unnatural within the string theory realm. It seems however worth inspecting thoroughly these theories based on the very fact that they lead to {\it reasonable} results.  That is, even for these presumably non-stringy setups we obtain what seem to be splendid checks of the AdS/CFT correspondence.

Section 6 is devoted to the analysis of the $\eta/s$ ratio in higher order Lovelock theories. The first part of that section presents analytic and numerical results for the minimum ratio in any dimension up to quartic order theory. These are obtained by considering the full stability and causality constraints. Our results are relatively smooth functions of the spacetime dimensionality.
In the second part we consider a limited analysis of arbitrary Lovelock theories in any dimension. Picking a specific curve through parameter space, the properties of which are carefully discussed in Appendix \ref{appendixB}, we show that there seems not to be a dimension independent viscosity bound. Within our framework, we argue that a strongly coupled ideal fluid could be achieved in the strict limit $d \to \infty$, when a correspondingly infinite number of Lovelock terms are taken into account. This is somehow reminiscent of the so-called {\it species problem}, though it is not clear to us whether there is a more rigorous way in which these two setups are related. Following a discussion and conclusions presented in section 7, we have a series of appendices where we derive a holographic formula for the parameters entering the $3$-point stress energy tensor correlations in arbitrary dimensions, and we present the details of the analysis of a particular curve through parameter space in an arbitrary Lovelock theory.

\section{Lovelock gravity}

Lovelock gravity is the most general second order gravity theory in higher-dimensional spacetimes, and it is free of ghosts when expanding about flat space \cite{Lovelock1971,Zumino1986}. The action can be written in terms of differential forms\footnote{We make use of the language of tensorial forms: out of the vierbein, $e^a$, and spin connection, $\omega_{~b}^a$, 1-forms, we construct the Riemann curvature, $R_{~b}^a$, and torsion, $T^a$, 2-forms:
\begin{equation*}
R_{~b}^a = d\,\omega_{~b}^a + \omega_{~c}^a \wedge \omega_{~b}^c =  \frac{1}{2} R_{~b\mu\nu}^{a}\; dx^{\mu} \wedge dx^{\nu} ~, \qquad T^a = d\,e^a + \omega_{~b}^a \wedge e^b ~.
\end{equation*}} as
\begin{equation}
\mathcal{I} = \sum_{k=0}^{K} {\frac{a_k}{d-2k}} \mathop\int \mathcal{R}^{(k)} ~,
\label{LLaction}
\end{equation}
where $a_k$ is a set of couplings with length dimensions $L^{2(k-1)}$, $L$ being some length scale related to the cosmological constant, and $d$ is the spacetime dimensionality. $K$ is a positive integer restricted to $K\leq \left[\frac{d-1}{2}\right]$. $\mathcal{R}^{(k)}$ is the exterior product of $k$ curvature 2-forms with the required number of vierbeins to construct a $d$-form, 
\begin{equation}
\mathcal{R}^{(k)} = \epsilon_{f_1 \cdots f_{d}}\; R^{f_1 f_2} \wedge \cdots \wedge R^{f_{2k-1} f_{2k}} \wedge e^{f_{2k+1} \cdots f_d} ~,
\end{equation}
where $e^{{f_1}\cdots{f_k}}$ is a short notation for $e^{f_1} \wedge \ldots \wedge e^{f_k}$. For convenience, we will also use from now on the expression $R^{f_1 \cdots f_{2k}} \equiv R^{f_1 f_2} \wedge \cdots \wedge R^{f_{2k-1} f_{2k}}$. The zeroth and first term in (\ref{LLaction}) correspond, respectively, to the cosmological term and the Einstein-Hilbert action. These are particular cases of the Lovelock theory. It is fairly easy to see that $a_0 = L^{-2}$ and $a_1 = 1$ correspond to the usual normalization of these terms.\footnote{We set, without any loss of generality, $16\pi (d-3)!\, G_N = 1$ to simplify our expressions ($G_N$ being the Newton constant in $d$ spacetime dimensions). In tensorial notation, thus, the cosmological constant has the customary negative value $2 \Lambda = - (d-1)(d-2)/L^2$.}

If we consider this action in the first order formalism, we have two equations of motion, for the connection 1-form and for the vierbein. If we vary the action with respect to the connection the resulting equation is proportional to the torsion, and so we can safely set it to zero as usual, allowing us to compare our results with those coming from the usual tensorial formalism based on the metric. The second equation of motion is obtained by varying the action with respect to the vierbein. It can be cast into the form
\begin{equation}
\mathcal{E}_a \equiv \epsilon_{a f_1 \cdots f_{d-1}}\; \mathcal{F}_{(1)}^{f_1 f_2} \wedge \cdots \wedge \mathcal{F}_{(K)}^{f_{2K-1} f_{2K}} \wedge e^{f_{2K+1} \cdots f_{d-1}} = 0 ~,
\end{equation}
where $\mathcal{F}_{(i)}^{a b} \equiv R^{a b} - \Lambda_i\, e^{a b}$, which makes manifest that, in principle, this theory admits $K$ constant curvature {\it vacuum} solutions,
\begin{equation}
\mathcal{F}_{(i)}^{a b} = R^{a b} - \Lambda_i\, e^{a b} = 0 ~.
\end{equation}
The $K$ different cosmological constants are the solutions of the $K$-th order polynomial
\begin{equation}
\Upsilon[\Lambda] \equiv \sum_{k=0}^{K} a_k\, \Lambda^k = a_K \prod_{i=1}^K \left( \Lambda - \Lambda_i\right) =0 ~,
\label{cc-algebraic}
\end{equation} 
each one corresponding to a `{\it vacuum}' which we may think of doing perturbations about. The theory will have degenerate behavior and possibly symmetry enhancement whenever two or more effective cosmological constants coincide. 

\subsection{AdS vacua and black hole solutions}

Here we are interested in AdS vacua and black brane solutions. These can be obtained by means of an ansatz of the form \cite{Wheeler1986,Wheeler1986a,Dehghani2009}\footnote{We can consider different functions in the timelike and radial directions but they are bound to be equal by the equations of motion for generic values of the couplings.}
\begin{equation}
ds^2 = - N_{\#}^2\, f(r)\, dt^2 + \frac{dr^2}{f(r)} + \frac{r^2}{L^2}\, d\mathbf x^2, \qquad \mathbf x = (x_2, \ldots, x_{d-1}) ~.
\label{bhansatz}
\end{equation}
A natural frame for the vierbein is then
\begin{equation}
e^0 = N_{\#}\, \sqrt{f(r)}\, dt ~, \qquad e^1 = \frac{1}{\sqrt{f(r)}}\, dr ~, \qquad e^a = \frac{r}{L}\, dx^a ~,
\label{vierbh}
\end{equation}
where $a = 2, \ldots, d-1$, and the only non-vanishing components of the spin connection read
\begin{equation}
\omega^0_{~1} = \frac{f'(r)}{2 \sqrt{f(r)}}\; e^0 ~, \qquad \omega^1_{~a}= - \frac{\sqrt{f(r)}}{r}\; e^a ~.
\label{spinbh}
\end{equation}
The Riemann 2-form is given by
\begin{eqnarray}
& & R^{01} = - \frac{ f''(r)}{2}\; e^0 \wedge e^1 ~, \quad\qquad R^{0a} = - \frac{f'(r)}{2 r}\; e^0 \wedge e^a ~, \nonumber \\ [1em]
& & R^{1a} = - \frac{f'(r)}{2 r}\; e^1 \wedge e^a ~, \quad\qquad R^{ab} = - \frac{f(r)}{r^2}\; e^a \wedge e^b ~.
\label{riemannbh}
\end{eqnarray}
If we insert these expressions into the equations of motion, we get after some manipulations
\begin{equation}
\left[ \frac{d~}{d\log r} + (d-1) \right]\, \left( \sum_{k=0}^{K} a_k\, g^k \right) = 0 ~,
\label{eqgg}
\end{equation}
where $g = - f/r^2$. This can be easily solved as
\begin{equation}
\Upsilon[g]\equiv\sum_{k=0}^{K} a_k\, g^k =\frac{1}{L^2} \left(\frac{r_+}{ r} \right)^{d-1}~,
\label{eqg}
\end{equation} 
where $r_+$ is an integration constant that determines the mass of the spacetime. If $r_+=0$, the equation is solved by the constant function $g(r)=\Lambda$, with $\Lambda$ a solution to \reef{cc-algebraic}, and the metric reduces to pure (A)dS. The full metric is fixed by choosing $N_{\#}^2=-(L^2\,\Lambda)^{-1}$, so that the speed of light in the boundary theory is set to one.

For $r_+>0$ the equation (\ref{eqg}) admits a non-trivial black hole solution with a horizon at $r=r_+$, where $g(r)$ vanishes. Notice that (\ref{eqg}) leads to $K$ different branches, each one associated to each of the cosmological constants (\ref{cc-algebraic}). Just one branch has a horizon there. This can be seen since the polynomial root $g=0$ has multiplicity one at $r=r_+$. In order to have higher multiplicity it would be necessary that $a_1$ vanishes, but $a_1=1$ in our treatment (it corresponds to the Einstein-Hilbert term). This unique branch of solutions with a horizon -- from now on the `relevant branch' -- is just a deformation of the usual Einstein-Hilbert AdS black hole solution and shares with it a lot of nice features. For instance, it has a well defined temperature
\begin{equation}
T = N_{\#} \frac{f'(r_+)}{4\pi} = \frac{(d-1)N_{\#}}{4\pi}\; \frac{r_+}{L^2} ~,
\end{equation}
as $r g'(r)= -(d-1)\,\frac{\Upsilon[g]}{\Upsilon'[g]}$, and its entropy still satisfies the usual area law \cite{Myers1988,Jacobson1993,Nojiri2001j,Cai2002,Kofinas2006,Cai2004}
\begin{equation}
s = \frac{r_+^{d-2}}{4} ~,
\end{equation}
where we have divided by the volume of the flat directions. The mass can be obtained as
\begin{equation}
m = \int T\,ds = \int_{0}^{r_+} T\,\frac{ds}{dr_+}\,dr_+ = \frac{(d-2)N_{\#}}{16\pi}\; \frac{r_+^{d-1}}{L^2} ~.
\end{equation}
This is general for any Lovelock black hole of the type considered here. 

A detailed study of black hole solutions in Lovelock theory is far from trivial (see, for instance, \cite{Garraffo2008}). For cubic (or higher order) Lovelock gravities, there are entire regions of the parameter space where there are no black holes at all even if there is a well defined AdS vacuum \cite{Boer2010,Camanho2010a}. This is a novel feature that does not take place in Gauss-Bonnet gravity. It has to do with the several branches mentioned above. The full solution is not valid whenever the relevant branch degenerates with any other one for a given radius. A careful analysis leads to the characterization of a non-compact subset of the space of Lovelock couplings that do not admit black hole solutions \cite{Camanho2010a}.

\section{Holographic dictionary}

Lovelock theory has a rich structure of AdS vacua. Correspondingly, the AdS/CFT correspondence tells us that there should be an analogously complex structure of dual CFTs. We know very little about these higher-dimensional CFTs. Indeed, at least for the case of supersymmetric CFTs, one naively expects to have at most six dimensional non-trivial unitary conformal field theories, based on the algebraic construction by Nahm \cite{Nahm1978}. For higher-dimensional CFTs, the general expectation is that they are not lagrangian theories and their constituent degrees of freedom are not gauge fields but possibly self-dual p-forms (in the case of even spacetime dimensions) \cite{Witten2007c}.  These theories should have a stress tensor and conformal symmetry will constrain their $2$- and $3$-point functions as we shall presently describe.

\subsection{Calculation of $C_T$}

The leading singularity of the $2$-point function for a general CFT in any number of dimensions is fully characterized by the central charge $C_T$ \cite{Osborn1994} as
\begin{equation}
\langle T_{ab}(\mathbf x)\, T_{cd}(\mathbf 0)\rangle=\frac{C_T}{\mathbf x^{2(d-1)}}\;\mathcal{I}_{ab,cd}(\mathbf x)~, 
\end{equation}
where 
\begin{equation}
\mathcal{I}_{ab,cd}(\mathbf x) = \frac12 \left( I_{ac}(\mathbf x)\, I_{bd}(\mathbf x) + I_{ad}(\mathbf x)\, I_{bc}(\mathbf x) - \frac1{d-1}\, \eta_{ab}\, \eta_{cd} \right) ~,
\end{equation}
and
\begin{equation}
I_{ab}(\mathbf x) = \eta_{ab} - 2\, \frac{x_a\, x_b}{\mathbf x^2} ~.
\end{equation}
This structure is completely dictated by conformal symmetry.

The holographic computation of $C_T$ was performed in \cite{Buchel2010a} for Gauss-Bonnet theory in various dimensions. If one considers a general Lovelock theory, the result is very similar as we shall see shortly. Since $C_T$ appears in the general $2$-point function of the stress tensor, it is sufficient to consider a particular set of components. Following \cite{Buchel2010a} we consider the correlator $\langle T_{xy}(\mathbf x)\,T_{xy}(\mathbf 0)\rangle$. According to the AdS/CFT dictionary, such a correlator can be computed by considering gravitational fluctuations around the AdS vacuum of the theory. In the present case, it is sufficient to take a metric fluctuation $\phi(r,\mathbf x) \equiv h_{x}^{~\,y}(r,\mathbf x)$ about empty AdS. Expanding the lagrangian  \reef{LLaction} to quadratic order in the fluctuation and evaluating it on-shell, we find
\begin{equation}
\mathcal{I}= \frac{1}{8 (-\Lambda)^{\frac{d-2}{2}} L_p^{d-2}}  \Upsilon'[\Lambda] \int{\,d{\bf x}\,dt\;r^{d} \left(\phi\,\partial_r\phi\right)} ~,
\end{equation}
where $L_p^{d-2}=8\pi G_N$ is the Planck length and %
\be
\Upsilon'[\Lambda]=\sum_{k=1}^{K}{k\,a_k\,\Lambda^{k-1}} ~.
\ee
In the notation of \cite{Buchel2010a} this expression is given by\footnote{We have the equivalence $\ff=-\Lambda L^2$.} $(1-2 \ff \lambda+\ldots)$. Indeed, the preceding formula exactly matches equation (3.12) in that reference with $\Upsilon'[\Lambda]$ corresponding to the specific case of Gauss-Bonnet. Accordingly, our calculation of $C_T$ proceeds in an analogous fashion and we find
\be
C_T=\frac{d}{d-2} \frac{\Gamma[d]}{\pi^{\frac{d-1}2} \Gamma\left[\frac{d-1}2\right]} \frac{\Upsilon'[\Lambda]}{(-\Lambda)^{d/2}} ~. 
\ee
If there is a dual conformal field theory, or class of theories which are holographically described by the Lovelock gravity lagrangian, then the above expression provides the value of the central charge for those theories. Unitarity then demands that $C_T$ must be positive, and this nicely translates into the gravity side to the condition $\Upsilon'[\Lambda]>0$. From the above, we see that if this is not so, the kinetic term for the graviton modes acquires a negative sign, and hence these modes are no longer unitary and the theory shows the kind of instabilty first found by Boulware and Deser (BD) \cite{Boulware1985a} in Gauss-Bonnet gravity. The positivity of $\Upsilon'[\Lambda]$ has been proven to be true for the relevant branch of solutions in any Lovelock theory \cite{Camanho2010a}. Thus, this branch is protected from BD instabilities and the corresponding dual theory is unitary. 

\subsection{Three-point function}

The form of the $3$-point function of the stress-tensor in a $(d-1)$-dimensional conformal field theory is highly constrained. In \cite{Osborn1994,Erdmenger1997}, it was shown that it can always be written in the form
\be
\langle T_{ab}(\mathbf x)\,T_{cd}(\mathbf y)\,T_{ef}(\mathbf z)\rangle=
\frac {\left( \mathcal A\, \mathcal I^{(1)}_{ab,cd,ef}+\mathcal B\, \mathcal I^{(2)}_{ab,cd,ef}+\mathcal C\, \mathcal I^{(3)}_{ab,cd,ef}\right)}{(|\mathbf x - \mathbf y|\,|\mathbf y - \mathbf z|\,|\mathbf z - \mathbf x|)^{d-1}}
\ee
where the form of the tensor structures $\mathcal I^{(i)}_{ab,cd,ef}$ will be irrelevant for us here. Energy conservation also implies a relation between the central charge $C_T$ appearing in the $2$-point function, and the parameters $\mathcal A,\mathcal B,\mathcal C$, namely
\be
C_T=\frac{ \pi^{\frac{d-1}{2}}}{\Gamma\left[\frac{d-1}2\right]}\,\frac{(d-2)(d+1)\mathcal{A}-2\mathcal{B}-4 d \, \mathcal{C}}{(d-1)(d+1)}~.
\ee
Since we have already computed $C_T$ in the previous section, we are left with two independent parameters to be calculated.

A convenient parameterization of the $3$-point function of the stress-tensor was introduced in \cite{Hofman2008}. One considers a localized insertion of the form $\int d\omega\, \epsilon_{jk}\,e^{-i\omega t}\, T^{jk}(\mathbf x)$, and measures the energy flux at lightlike future infinity along a certain direction $\mathbf n$. The final answer for the energy flux $\mathcal E(\mathbf n)$ takes the form \cite{Camanho2010,Buchel2010a}
\be
\langle \mathcal E(\mathbf n)\rangle=\frac{E}{\Omega_{d-3}}\left[1+t_2 \left(\frac{n_i n_j \epsilon^*_{ik}\epsilon_{jk}}{\epsilon^*_{ik}\epsilon_{ik}}-\frac 1{d-2}\right)
+t_4 \left(\frac{|n_i n_j\epsilon_{ij}|^2}{\epsilon^*_{ik}\epsilon_{ik}}-\frac 2{d(d-2)}\right)\right]~,
\ee
with $E$ the total energy of the insertion, and $\Omega_{d-3}$ the volume of a unit $(d-3)$-sphere.

The existence of minus signs in the above expression leads to interesting constraints on the parameters $t_2$ and $t_4$, by demanding that the measured energy flux should be positive for any direction $\mathbf n$ and polarization $\epsilon_{ij}$. In spite of the fact that the positivity of the energy flux is not self-evident, there is a field theoretic argument supporting this claim \cite{Hofman2009}. Furthermore, interestingly enough, it was recently argued that this condition is equivalent to unitarity of the corresponding CFT \cite{Kulaxizi2010a}. By picking a particular direction and considering polarizations with different rotational transformation properties with respect to it, one finds three different constraints,
\bea
{\rm Tensor:} & & \qquad 1-\frac{1}{d-2}t_2-\frac 2{d(d-2)} t_4 \geq 0~, \nonumber\\[0.4em]
{\rm Vector:} & & \qquad 1+\frac{d-4}{2(d-2)}t_2-\frac 2{d(d-2)} t_4 \geq 0~, \label{posenergy}\\[0.4em]
{\rm Scalar:} & & \qquad 1+\frac{d-4}{d-2}t_2+\frac {d(d-3)-2}{d(d-2)} t_4 \geq 0~. \nonumber 
\eea

For conformal field theories with a weakly curved gravitational dual, it is possible to compute $t_2$ and $t_4$ holographically. This was first done in detail for Gauss-Bonnet theory in various dimensions \cite{Buchel2010a} and later applied to quasi-topological gravity in \cite{Myers2010d}. 
The calculation proceeds by considering the vacuum AdS solution perturbed by a shockwave, which corresponds holographically to a $T_{--}$ insertion. By adding a transverse metric fluctuation, one reads off the interaction vertex from the action, and from that one obtains $t_2$ and $t_4$.

Shockwave backgrounds were also considered by Hofman \cite{Hofman2009} in the context of five-dimensional Gauss-Bonnet theory. There it was found that in the presence of the shockwave there is the possibility for causality violation in the dual field theory. This places bounds on the parameters of the theory, which precisely match those portrayed in (\ref{posenergy}). This shockwave calculation was extended to arbitrary higher-dimensional Gauss-Bonnet gravity in \cite{Camanho2010}, and the corresponding constraints were found. In \cite{Buchel2010a}, $t_2$ and $t_4$ were computed explicitly for these theories and the bounds (\ref{posenergy}) perfectly agree with the constraints obtained in \cite{Camanho2010}.

We would now like to generalize this story to higher Lovelock theories. We consider, along the lines of \cite{Camanho2010}, a helicity two perturbation $\phi(u,v,r)$ in the shockwave background 
\begin{equation}
ds^2_{{\rm AdS},sw}=\frac{N_{\#}^2\,L^2}{z^2}\left(-du\,dv+dx^i\,dx^i+2\,\epsilon\,\phi\,dx^2\,dx^3+dz^2\right)+f(u)\,\varpi(x^{a},z)\,du^2~,
\end{equation}
where $z=L^2/r$ is the Poincar\'e coordinate and $u,v=x^0\pm x^{d-1}$ are light-cone coordinates. This amounts to choosing just one non-vanishing component of the polarization tensor, $\epsilon_{23}\neq0$. Leading contributions (in the high momentum limit) to the equations of motion come from the exterior derivative of the perturbation of the spin connection.\footnote{The $3$-point function will be determined by these kinds of terms; actually, terms in $\partial_v^2\phi$ since the vertex has to be of the form $\phi\,\partial_v^2\phi\,\partial_i\partial_j\varpi$, as $\varpi\sim h_{uu}$.} We get,
\begin{eqnarray}
& & d(\delta \omega^{02})  \approx \frac{\epsilon\; z^2}{L^2 N^2_{\#}}\, \left[ \partial_v^2 \phi\; e^1 \wedge e^2 + \left( \partial_u \partial_v \phi + \frac{z^2}{L^2 N^2_{\#}}\, f(u) \,\varpi(x^a, z) \, \partial_v^2 \phi \right)\, e^0 \wedge e^2 \right]  ~, \label{dw02} \nonumber \\ [0,7em]
& & d(\delta \omega^{12}) \approx \frac{\epsilon\; z^2}{L^2 N^2_{\#}}\, \left[ \left( \partial_u \partial_v \phi + \frac{z^2}{L^2 N^2_{\#}}\, f(u) \,\varpi(x^a, z) \, \partial_v^2 \phi \right)\, e^1\wedge e^2 + \left( \cdots \right)\, e^0\wedge e^2 \right] ~, \label{dw12} \nonumber
\end{eqnarray}
the ellipsis being used in the second expression since the corresponding term does not contribute to the equations of motion. The components with index 3 instead of 2 are the only remaining non-vanishing ones, and they are obtained just by changing\footnote{$dx^2\rightarrow -dx^3$, $dx^3\rightarrow dx^2$, $\phi\rightarrow-\phi$ is a symmetry of the background as well as of the vierbein basis.} $\phi\rightarrow- \phi$. The other ingredient we need is the curvature 2-form of the background metric, that can be written as
\begin{equation}
R^{ab} = \Lambda (e^a\wedge e^b + f(u) X^{ab}) ~,
\label{RabXab}
\end{equation}
where $\Lambda=-\frac{1}{L^2 N_{\#}^2}$ and $X^{ab}$ is an antisymmetric 2-form accounting for the contribution of the shock wave. The relevant component here is 
\begin{eqnarray}
& & X^{1a} = \frac{z^2}{L^2 N^2_{\#}}\, \left[\left(2\varpi+z\partial_z \varpi-z^2 \partial_a^2\varpi\right)\; e^0\wedge e^a+\left(\cdots\right)\; e^0\wedge e^z+\left(\cdots\right)\; e^0\wedge e^b\right]  ~,  \nonumber \\ [0.7em]
& & X^{1(d-1)} = -\left(\frac{z}{L N_{\#}}\right)^2\, \left[(3z\partial_z
\varpi+z^2 \partial_z^2\varpi)\; e^0 \wedge
e^{d-1}+\left(\cdots\right)\; e^{0}\wedge e^{a}\right] ~, \nonumber
\end{eqnarray}
with $a,b\neq 0,1,d-1$, and $b\neq a$. The relevant equation of motion is given by $\delta \mathcal{E}_3\wedge e^3 =0$, where
\begin{eqnarray}
& & \delta \mathcal{E}_3\wedge e^3 = \sum_{k=1}^{K}{}k\,a_k\; \epsilon_{3 f_1 \cdots f_{d-1}} d(\delta \omega^{f_1 f_2})\wedge R^{f_3 \cdots f_{2k}}\wedge e^{f_{2k+1}\cdots\, f_{d-1} \,3} \nonumber \\ [0.9em]
& & \quad = 4(-1)^{d-1} \frac{\epsilon\; z^2}{L^2 N^2_{\#}}\, \sum_{k=1}^{K}{}k\,a_k\,\Lambda^{k-1}\left[(d-3)!\left( \partial_u \partial_v \phi + \frac{z^2}{L^2 N^2_{\#}}\, f(u) \,\varpi \, \partial_v^2 \phi \right) \right.\nonumber\\ [0.9em]
& & \qquad - \left. (k-1) (d-5)!\; \frac{z^2}{L^2 N^2_{\#}}\, f(u)\left(-4\varpi-2z\partial_z\varpi+z^2(\partial_2^2\varpi+\partial_3^2\varpi)\right)\partial_v^2\phi\right] ~.
\label{eq3}
\end{eqnarray} 
We made use of the equation of motion for the shock wave profile
\begin{equation}
2(d-3)\varpi+(d-6)z\partial_z \varpi-z^2(\partial_a\partial^a\varpi+\partial_z^2 \varpi)=0~,\qquad a=2,\ldots d-2\; .
\label{SWeq}
\end{equation}
This equation admits several solutions of the type
\begin{equation}
\varpi =\alpha_0 \frac{z^\alpha}{\left(z^2+(x^2-x^2_0)^2+\cdots+(x^n-x^n_0)\right)^\beta} ~,
\end{equation}
where $n$ is the number of transverse coordinates. The relevant solution for our discussion here is the one given by $\alpha=d-3$ and $\beta=d-2$. This shockwave profile has been argued \cite{Hofman2008} to be the dual field configuration to $\mathcal{E}(\mathbf n)$ provided $x_0^i=\frac{n^i}{1+n^{d-2}}$ and $f(u)=\delta(u)$. 

From (\ref{eq3}), we shall focus on those terms proportional to $\partial_{v}^2\phi$. The shockwave interaction term thus gives the following contribution to the equations of motion,
\begin{eqnarray}
\left[\Upsilon'[\Lambda]\, \varpi+\frac{\Lambda\,\Upsilon''[\Lambda]\,}{(d-3)(d-4)}\left(4\varpi+2z\partial_z \varpi-z^2(\partial_2^2\varpi+\partial_3^2\varpi)\right)\right]\partial_v^2\phi~.
\end{eqnarray}
There are extra non vanishing components of the form
\begin{eqnarray*}
& & \delta \mathcal{E}_{i}\wedge e^2\sim \partial_3\partial_i\varpi \partial^2_{v}\phi ~, \\ [1em]
& & \delta \mathcal{E}_{d-1}\wedge e^2\sim \frac1{x^{d-1}}\left(\partial_3\varpi +x^{d-1} \partial_3\partial_{d-1}\varpi\right)\partial^2_{v}\phi ~, \\ [1em]
& & \delta \mathcal{E}_{i\neq2,3}\wedge e^{j\neq2,3}\sim \partial_2\partial_3\varpi \partial^2_{v}\phi \delta_i^j ~.
\end{eqnarray*}
However, these are irrelevant for computing the three-point function we are interested in. The tensor channel mixes with other modes but they would only affect other correlators irrelevant for our discussion.

The $3$-point function follows from evaluating the effective action for the field $\phi$ on-shell, on a particular solution which depends on all coordinates, including $x_2,x_3$ \cite{Buchel2010a}. The cubic interaction vertex of $\phi$ with the shockwave appearing in the action will be essentially the one in the equation of motion determined above. Up to an overall factor, the cubic vertex is then
\be
S^{(3)}\sim C_T \int d^d x \sqrt{-g} \,\phi\, \partial^2_v \phi\, \varpi\,\left(1- \frac{\Lambda\,\Upsilon''(\Lambda)}{\Upsilon'(\Lambda)}\frac{T_2}{(d-3)(d-4)}\right)~,
\ee
where 
\be
T_2=\frac{z^2(\partial_2^2 \varpi+\partial_3^2 \varpi)-2 z \partial_z \varpi-4 \varpi}{\varpi}~.
\ee
This is nothing but the same $T_2$ appearing in \cite{Buchel2010a}. Indeed, following that paper the relevant graviton profile is
\begin{equation}
\phi(u=0,v,x^a,z)\sim e^{-iEv}\,\delta^{d-3}(x^a)\,\delta(z-1)~,
\end{equation}
so that we need to impose $x^a=0$ and $z=1$ yielding
\be
T_2=2(d-1)(d-2)\left(\frac{n_2^2+n_3^2}{2}-\frac1{d-2}\right)~,
\ee
and we therefore read off
\be
t_2=-\frac{2(d-1)(d-2)}{(d-3)(d-4)} \frac{\Lambda \Upsilon''[\Lambda]}{\Upsilon'[\Lambda]}~, \quad\qquad t_4=0~. \label{t2t4}
\ee
This expression reproduces previous results in Gauss-Bonnet gravity \cite{Camanho2010,Buchel2010a}, and is exactly the same as conjectured by de Boer, Kulaxizi and Parnachev in \cite{Boer2010}. This completely proves the equivalence between the positivity of energy constraints and those coming from causality in the boundary theory (see next Section). Using these results together with the expression for $C_T$, we find expressions for the usual $3$-point function parameters $\mathcal A,\mathcal B, \mathcal C$ in Appendix \ref{3point}.

\section{Shear viscosity and finite temperature instabilities}

It has been shown in \cite{Brustein2009a,Shu2010} that the shear viscosity to entropy ratio of a generic Lovelock theory is given by
\be
\frac{\eta}s = \frac{1}{4\pi}\left(1-\frac{\lambda}{\lambda_c}\right) ~,
\label{etaeq}
\ee
where $\lambda=a_2/L^2$, and we have defined the critical value $\lambda_c$ where the ratio vanishes
\be
\lambda_c \equiv \frac{d-3}{2 (d-1)} ~.
\label{lambdac}
\ee
It is interesting to stress that the ratio only depends on the quadratic Lovelock coupling, $a_2$. Generically one is not free to take an arbitrary $\lambda$. There are causality constraints which restrict the range of values where the theory is well-defined \cite{Brigante2008,Brigante2008a}. These restrictions match the restrictions on $t_2$ and $t_4$, and these necessarily involve all the parameters in the theory. The maximum value $\lambda_{(0)}$ that may be achieved depends on the theory in question. In this way it has been found \cite{Boer2010,Camanho2010a} that causality constraints alone cannot prevent one from reaching (and even overpassing) $\eta/s=0$. In this section we show that as one approaches small values of $\eta/s$, the black hole becomes unstable and the linear approximation breaks down. This places an effective set of constraints in the parameters of the theory, saving the ratio above from ever becoming too small. But first, let us briefly recall how to derive the constraints by demanding causality.

\subsection{Graviton potentials and causality}
\label{causal}

We start off with the black hole background \reef{bhansatz}, and add metric fluctuations $h_{ab}$ of frequency $\omega$ and momentum $q$ in a fixed direction. The fluctuations split into three channels according to their polarization relative to the momentum, namely the tensor, shear and sound channels \cite{Policastro2002}  (or equivalently helicity/spin two, one or zero). The equations of motion for these dynamical degrees of freedom, $\phi_i(r)$, the subindex $i=0, 1, 2$ indicating the corresponding helicity, can be recast as Schr\"odinger type equations \cite{Takahashi2010}. In the large momentum limit they reduce to a very simple form\footnote{In the notation of \cite{Takahashi2010}, we must identify $\gamma_i \equiv \gamma = q^2 L^2$, and our potentials are related to theirs, $U_i \to - V_i/\gamma\,\Lambda$, as $\gamma \to \infty$.}
\begin{equation}
\label{Scheq}
-\hbar^2\, \partial^2_y \Psi_i+U_i(y)\,\Psi_i=\alpha^2\,\Psi_i~,\qquad\qquad \hbar\equiv\frac{1}{q}\rightarrow 0~,
\end{equation}
where $\alpha = \omega^2/q^2$, $y$ is a dimensionless tortoise coordinate defined as $dy/dr=\sqrt{-\Lambda}/f(r)$, and $\Psi_i(y) = B_i(y)\,\phi_i(y)$, where $B_i(y) = B_i(g(y))$ are functions of the metric whose specific expression can be found in \cite{Takahashi2010}. In order for the linear analysis to be valid we need a non-vanishing function $B_i(y)$; we shall comment on this later on. The effective potentials $U_i$, can be determined as the speed of large momentum gravitons in constant $y$ slices,
\begin{equation}
U_i(y) = \begin{cases} c_i^2 (y) & \qquad y < 0 ~, \cr
                      + \infty & \qquad y = 0 ~,
        \end{cases} 
\end{equation}  
$y=0$ being the boundary of the spacetime.

For a generic Lovelock theory, in terms of the original radial variable, one finds for the tensor, shear and sound channels, respectively \cite{Camanho2010a}:
\bea
c_2^2(r) & = & \frac{g/\Lambda}{(d-4)} \frac{\mathcal{C}^{(2)}_d[g,r]}{\mathcal{C}^{(1)}_d[g,r]} ~, \nonumber\\ [0.6em]
c_1^2(r) & = & \frac{g/\Lambda}{(d-3)} \frac{\mathcal{C}^{(1)}_d[g,r]}{\mathcal{C}^{(0)}_d[g,r]} ~, \label{pots}\\ [0.6em]
c_0^2(r) & = & \frac{g/\Lambda}{(d-2)} \left(\frac{2\,\mathcal{C}^{(1)}_d[g,r]}{\mathcal{C}^{(0)}_d[g,r]}-\frac{\mathcal{C}^{(2)}_d[g,r]}{\mathcal{C}^{(1)}_d[g,r]}\right) ~, \nonumber
\eea
where we have defined functionals, $\mathcal{C}^{(k)}_d[g,r]$, involving up to $k$th-order derivatives of $g$:
\begin{eqnarray}
{C}^{(0)}_d[g,r] & = & \Upsilon' [g]~, \\ [0.9em]
{C}^{(1)}_d[g,r] & = & \left(r \frac{d}{dr}+(d-3)\right)\Upsilon' [g] ~,\\ [0.6em]
{C}^{(2)}_d[g,r] & = & \left(r \frac{d}{dr}+(d-3)\right)\left(r \frac{d}{dr}+(d-4)\right)\Upsilon' [g]~.
\end{eqnarray}
Notice now that we can make a change of variables $(r,g)\rightarrow (x\equiv\log L^2\Upsilon,F\equiv\log L^2\Upsilon')$ so that $r\partial_r=-(d-1)\partial_x$ (as $r$ and $\Upsilon$ are related by (\ref{eqg})) yielding for the potentials
\begin{eqnarray}
c_2^2(x) & = & \frac{(d-1)\,g}{(d-4)\,\Lambda}\frac{\left(\frac{d-3}{d-1}-F'(x)\right)\left(\frac{d-4}{d-1}-F'(x)\right)+F''(x)}{\left(\frac{d-3}{d-1}-F'(x)\right)} ~, \nonumber\\[0.6em]
c_1^2(x) & = & \frac{(d-1)\,g}{(d-3)\,\Lambda}\left(\frac{d-3}{d-1}-F'(x)\right) ~, \label{potentialsF}\\ [0.6em]
c_0^2(x) & = & \frac{(d-1)\,g}{(d-2)\,\Lambda}\frac{\left(\frac{d-3}{d-1}-F'(x)\right)\left(\frac{d-2}{d-1}-F'(x)\right)-F''(x)}{\left(\frac{d-3}{d-1}-F'(x)\right)} ~.\nonumber 
\end{eqnarray}
In order to avert causality violation, we must demand these effective potentials to be always smaller than one \cite{Brigante2008,Brigante2008a}. In particular, at the boundary we have $c_i^2=1$ and there we must demand\footnote{Recall that the boundary corresponds to $x_b=-\infty$ and the horizon to $x_h=0$.} $\partial_x c_i^2 \leq 0$, as $x \to -\infty$. Using $\partial_x g=\Upsilon[g]/\Upsilon'[g]$, we get the following constraints:
\begin{eqnarray}
{\rm Tensor:}\qquad & & \Upsilon'[\Lambda]+\frac{2 (d-1)}{(d-3)(d-4)}\,\Lambda\Upsilon''[\Lambda]\geq 0~.\nonumber\\[0.7em]
{\rm Vector:}\qquad & & \Upsilon'[\Lambda]-\frac{ (d-1)}{(d-3)}\,\Lambda\Upsilon''[\Lambda]\geq 0~.\\[0.7em]
{\rm Scalar:}\qquad & & \Upsilon'[\Lambda]-\frac{2 (d-1)}{(d-3)}\,\Lambda\Upsilon''[\Lambda]\geq 0~.\nonumber
\label{genconstraints}
\end{eqnarray}
These can be rewritten in terms of the dual CFT parameters, using the expressions for $t_2, t_4$ in \reef{t2t4}. The constraints become:
\bea
{\rm Tensor:} & & \qquad  1-\frac{1}{d-2}\, t_2 \geq 0 ~. \nonumber \\ [0.7em]
{\rm Vector:} & & \qquad  1+\frac{d-4}{2(d-2)}\, t_2 \geq 0 ~. \\ [0.7em]
{\rm Scalar:} & & \qquad  1+\frac{d-4}{d-2}\, t_2 \geq 0 ~. \nonumber
\eea
These precisely match the constraints (\ref{posenergy}) coming from positivity of energy in the dual conformal field theory (with $t_4=0$).

\subsection{Plasma instabilities}

It has been found in \cite{Dotti2005,Gleiser2005,Buchel2010a} that, for certain values of the Gauss-Bonnet coupling, some effective potentials might develop negative values close to the horizon. This has been previously argued to indicate an instability of the plasma \cite{Myers2007}. In eq.(\ref{Scheq}) we see that the role of $\hbar$ is played by $1/q$. By taking sufficiently large spatial momentum (small $\hbar$), we can make an infinitesimally small (negative energy) well to support a negative energy state in the effective Schr\"odinger problem. Going back to the original fields, this translates into an exponentially growing and therefore unstable mode \cite{Myers2007}. 

We now show that the same phenomenon occurs in a general Lovelock theory\footnote{Some work in this direction has been done recently in \cite{Takahashi2009,Takahashi2009a}.}. A negative potential well develops whenever the slope of the effective potential at the horizon is negative, and so we must require $\partial_x c_i^2\leq 0$ there. At the horizon we have $g=0$, and it is straightforward to make use of $\Upsilon(g)=L^{-2}e^{x}$ to find the derivatives of $g(x)$ at the horizon, {\it e.g.}, $\partial_x g|_{g=0} = \left(\Upsilon[g]/\Upsilon'[g]\right)|_{g=0} = a_0/a_1 = L^{-2}$. In this way, we can relate the values of the derivatives of $F(x)$ at $x=0$ (the horizon) with the coefficients of the polynomial $\Upsilon(g)$. In particular,
\begin{equation}
F'(0)=2\lambda ~, \qquad \qquad F''(0)= 2(\mu +\lambda(1-4\lambda)) ~,
\end{equation}
where, again, $\lambda=a_2/L^2$, and we have defined $\mu = 3 a_3/L^4$. Using these results, we expand the graviton effective potentials close to the horizon to get
\begin{eqnarray}
c_2^2(x) & \approx & \frac{d-1}{d-4}\frac{x}{L^2\,\Lambda}\frac{\left(\frac{d-3}{d-1}-2\lambda\right)\left(\frac{d-4}{d-1}-2\lambda\right)+2(\mu +\lambda(1-4\lambda))}{\left(\frac{d-3}{d-1}-2\lambda\right)}+\mathcal O\left(x^2\right) ~, \nonumber\\[0.7em]
c_1^2(x) & \approx & \frac{d-1}{d-3}\frac{x}{L^2\,\Lambda}\left(\frac{d-3}{d-1}-2\lambda\right)+\mathcal O\left(x^2\right) ~, \label{horexp}\\[0.7em]
c_0^2(x) & \approx & \frac{d-1}{d-2}\frac{x}{L^2\,\Lambda}\frac{\left(\frac{d-3}{d-1}-2\lambda\right)\left(\frac{d-2}{d-1}-2\lambda\right)-2(\mu +\lambda(1-4\lambda))}{\left(\frac{d-3}{d-1}-2\lambda\right)}+\mathcal O\left(x^2\right)\nonumber ~.
\end{eqnarray}
Stability requires that the potentials are positive. This gives rise to the following constraints for the Lovelock couplings:
\begin{eqnarray}
& &(d-3)-2 \lambda (d-1) > 0 \label{inscons1} ~, \\ [0.7em]
& & (d-3)(d-4)-2\lambda (d-1)(d-6)-4\lambda^2 (d-1)^2+2\mu (d-1)^2 \geq 0 \label{inscons2} ~, \\ [0.7em]
& & (d-2)(d-3)-6\lambda (d-1)(d-2)+12\lambda^2 (d-1)^2-2\mu (d-1)^2 \geq 0 \label{inscons0} ~.
\end{eqnarray}
This is one of our main results. These inequalities represent the constraints from the  shear, tensor and sound channels respectively. They are required to hold, otherwise the black brane solution is unstable. For $\mu=0$, these results match those for Gauss-Bonnet gravity derived in \cite{Buchel2010a}. It is important to notice that, in spite of considering a completely general Lovelock theory with as many terms as we wish, these stability constraints involve just the lowest two Lovelock couplings, $\lambda$ and $\mu$. 

The first constraint immediately rules out the region of the parameter space where the shear viscosity to entropy ratio becomes negative, since $\lambda < \lambda_c$. It is interesting to point out that the shear channel stability condition is also needed to ensure the validity of the linear analysis \cite{Takahashi2009}, as long as all the $B_i(r)$ functions are seen to be proportional to some power of the shear potential. Then, we can also relate the would be negative values of the shear viscosity -- that would lead to a manifest instability given its role as a damping coefficient in the sound channel (a negative value then amplifies the sound mode) --, to the accompanying break down of the linear analysis (at the horizon). In addition to this, the remaining constraints define a new allowed {\it stability wedge} in the parameter space:
\begin{eqnarray}
& & \mu \,\geq\, - \frac{(d-3)(d-4)-2\lambda (d-1)(d-6)-4\lambda^2 (d-1)^2}{2(d-1)^2} ~, \label{stabwedge1} \\ [0.7em]
& & \mu \,\leq\, \frac{(d-2)(d-3)-6\lambda (d-1)(d-2)+12\lambda^2 (d-1)^2}{2(d-1)^2} ~,
\label{stabwedge2}
\end{eqnarray}
with apex at
\begin{equation}
\lambda = \lambda_c = \frac{d-3}{2(d-1)} ~, \qquad \qquad \mu=\frac{(d-3)(d-5)}{2(d-1)^2} ~,
\end{equation}  
or, equivalently, on the intersection of the $\eta/s=0$ line with $\mu = \lambda (4\lambda - 1)$.
Then we can approach, from the inside of this stable region, a unique point (in the cubic case; a (hyper-)line in higher order cases) where $\eta/s=0$, and this is exactly at the apex of such region, where the stability constraints coming from each helicity meet.

For $d=7$, the apex coincides with the point of maximal symmetry $(\lambda,\mu)=(1/3,1/9)$, where the polynomial $\Upsilon[g]$ has a single maximally degenerate root. This is also the Chern-Simons point of the $d=7$ Lovelock theory.\footnote{The Chern-Simons point is the maximally symmetric point of a $K$-th order Lovelock theory in $d=2K+1$ dimensions, where it actually becomes a Chern-Simons theory for the AdS$_d$ group (see, for example, \cite{Zanelli2005}).} We plot the different regions in the space of parameters in figure \ref{StableCausal}.\!\!\!
\FIGURE{
\centering
		\includegraphics[width=123mm]{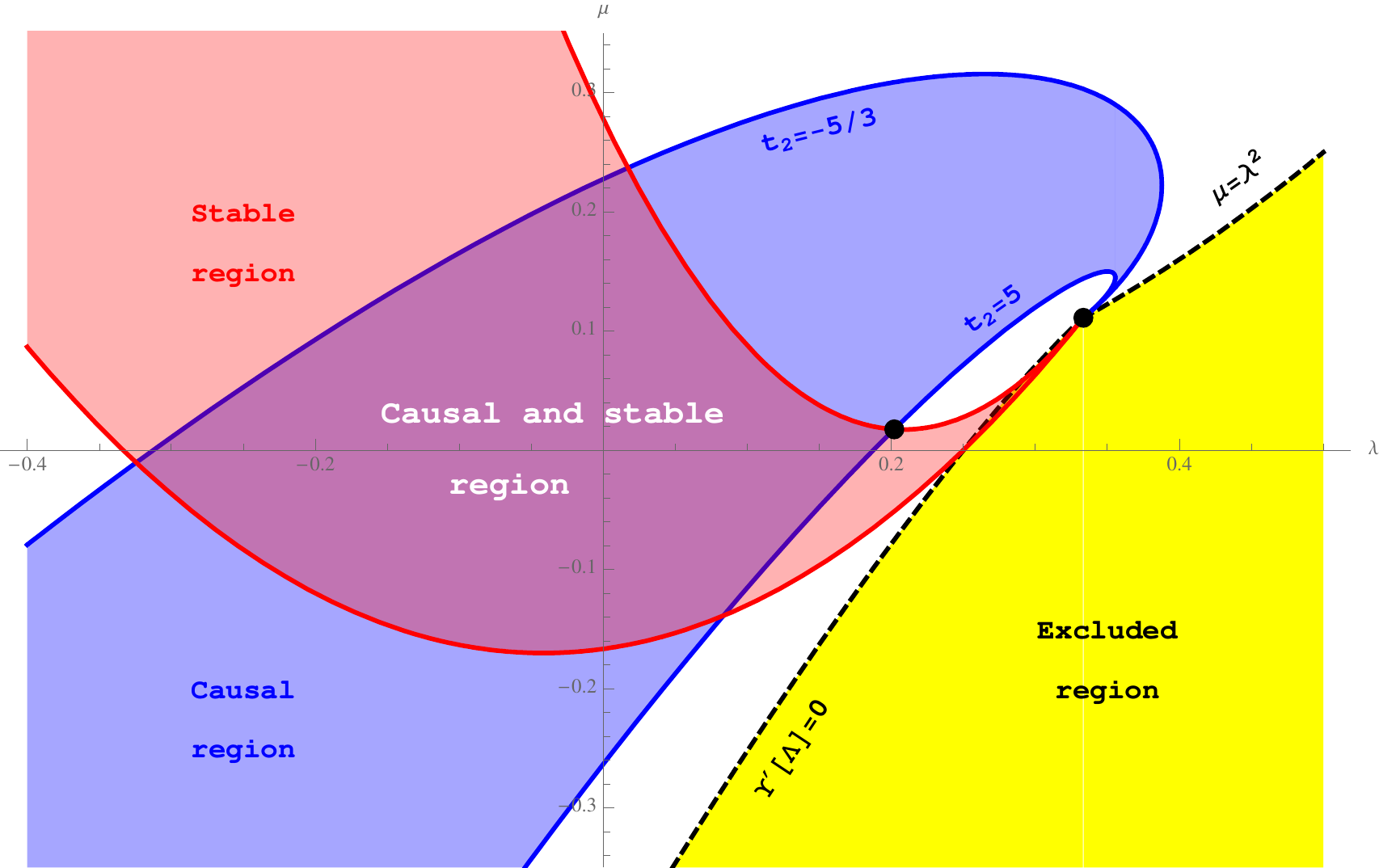}
  \caption{The allowed region by causality and stability for cubic Lovelock theory in $d=7$. The black points are the maximally degenerated point $(1/3,1/9)$ (also the Chern-Simons point in this case) and the intersection of the helicity zero stability constraint and the helicity two causality constraint $\sim(0.20,0.017)$, that gives the lowest value for $\eta/s$.}
{\label{StableCausal}}}
The minimum value for $\eta/s$ in this case is attained when the upper (helicity zero) stability curve intersects the lower (helicity two) causality curve, which happens at
\begin{equation}
\lambda \simeq 0.202042 ~, \qquad \qquad \mu \simeq 0.0175986 ~,
\end{equation}
giving a minimum value
\be
\frac{\eta}s \simeq 0.393874 \times \frac{1}{4\pi} < 0.4375 \times \frac{1}{4\pi} ~,
\ee
where the latter one is the correction to the KSS bound coming from GB alone (this corresponds to the value of $\lambda$ at which the lower blue curve intersects the axis, $\lambda_{\rm GB} = 3/16$). This seems to be the end of the story in $d=7$, at least in the context of Lovelock gravities. The existence of plasma instabilities (they originate in the behavior of the gravitational potential close to the horizon) sets a definite (positive) lower bound on cubic Lovelock theory. However, as we will see, the situation becomes more involved in higher dimensions, where the restrictions discussed so far are not enough to prevent $\eta/s$ from becoming zero.

\section{Bulk causality and stability}

\subsection{The cubic theory in higher dimensions}

Our results imply that the $\eta/s$ ratio can never be too small for cubic Lovelock theory in $d=7$ by a combination of two constraints -- preservation of causality, holographically dual to positivity of energy; and stability of the black brane solution which can be broadly identified with the stability of the thermal plasma. Of course the obvious question is whether similar results are valid in higher order theories. As it turns out, these two constraints are indeed enough, but not in the simple fashion we have described so far -- we must require causality and stability everywhere in the bulk, and not just from a simple near boundary or near horizon analysis.

\FIGURE{\includegraphics[width=0.5\textwidth]{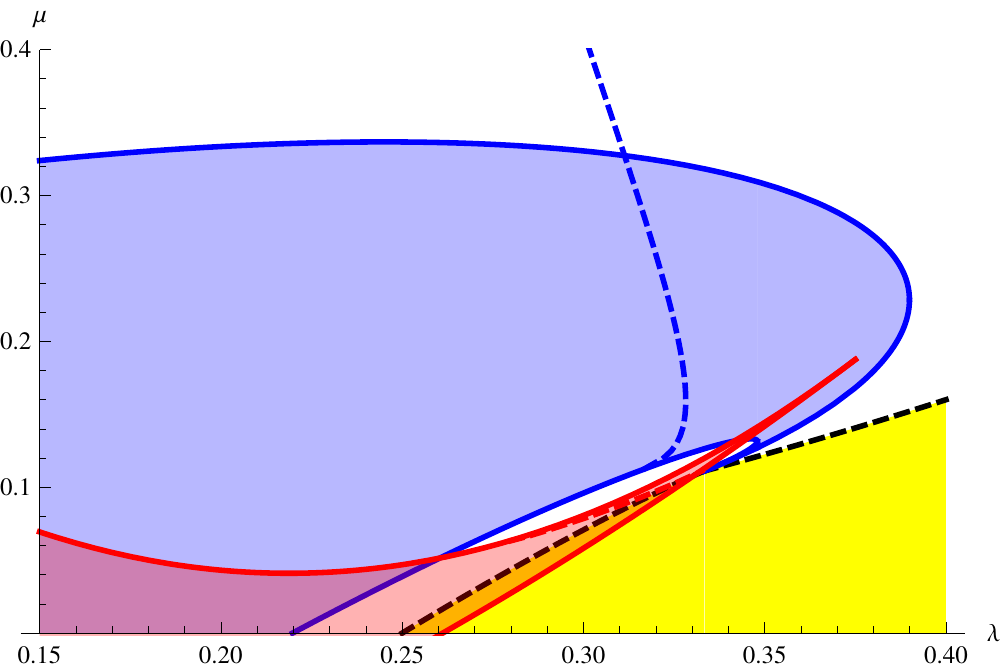}\includegraphics[width=0.5\textwidth]{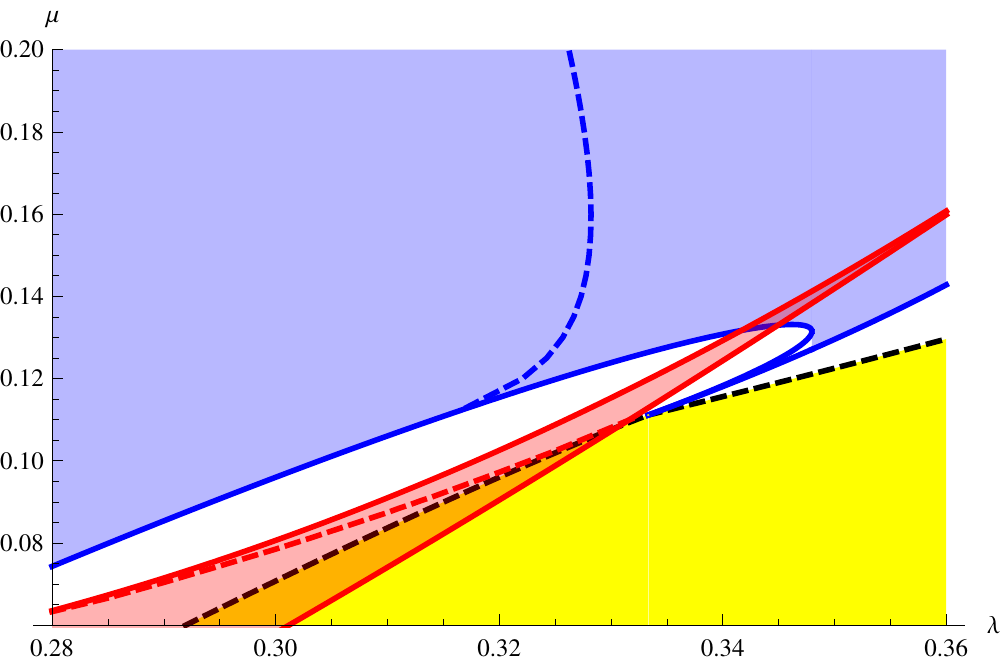}\caption{Causality and stability regions in $d=9$ cubic Lovelock theory (and zoom). The thick lines correspond to causality at the boundary and stability at the horizon, respectively in blue and red, while the dashed lines correspond to causality (in the tensor channel) and stability (in the sound channel) in the full geometry.}
{\label{StableCausal-9d}}}
\FIGURE{\includegraphics[width=0.5\textwidth]{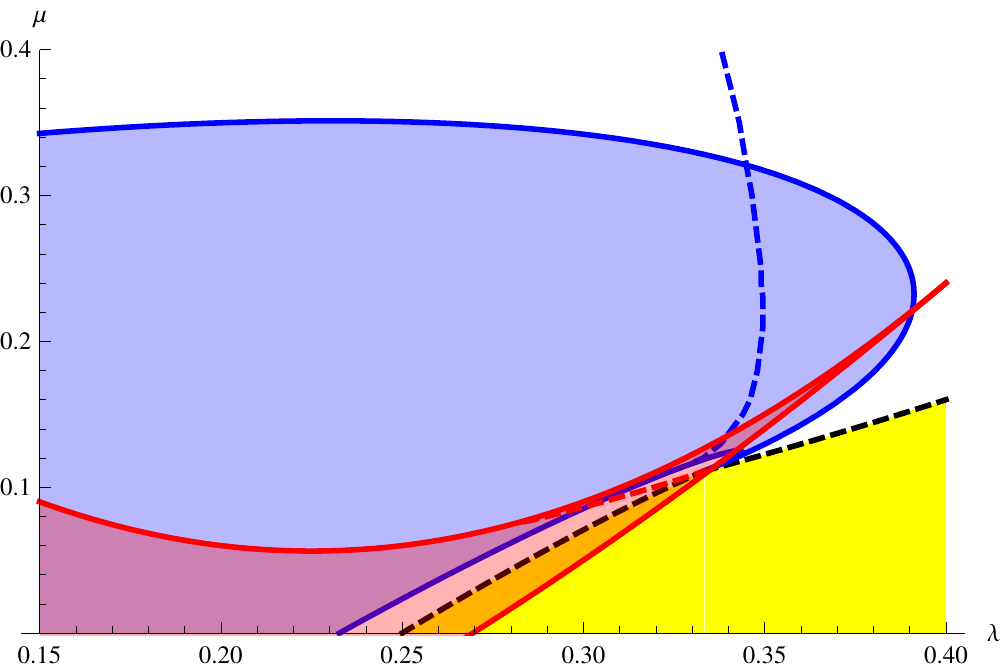}\includegraphics[width=0.5\textwidth]{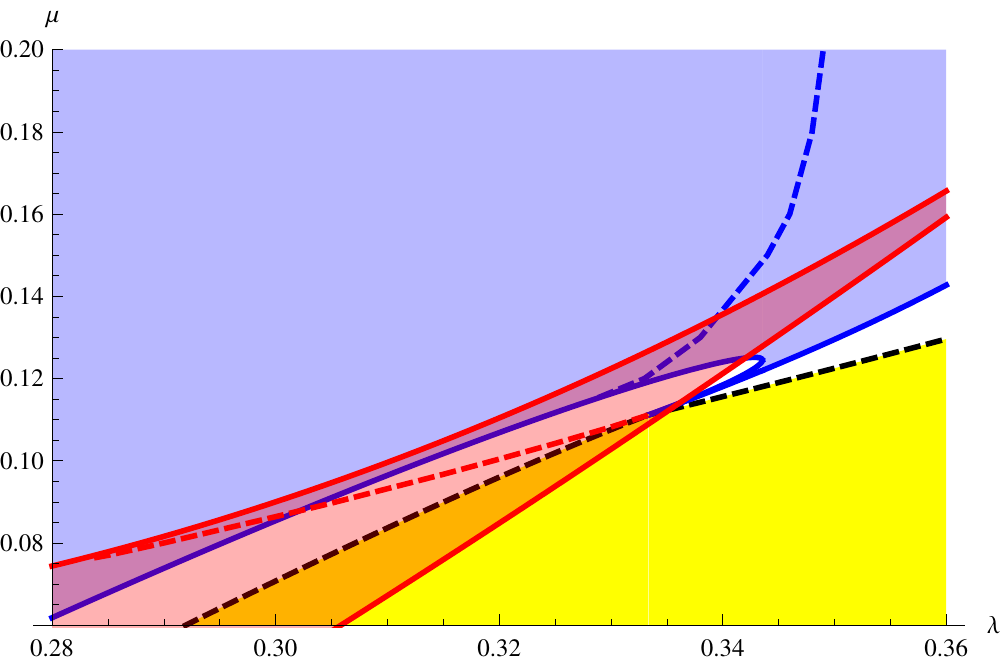}\caption{Causality and stability regions in $d=11$ cubic Lovelock theory (and zoom). The thick lines correspond to causality at the boundary and stability at the horizon, respectively in blue and red, while the dashed lines correspond to causality (in the tensor channel) and stability (in the sound channel) in the full geometry.}
{\label{StableCausal-11d}}}

This can be seen already in the simplest case of cubic Lovelock theory in higher dimensions. In figures \ref{StableCausal-9d} and \ref{StableCausal-11d} we show the relevant part of the causal and stable regions for $d = 9$ and $d = 11$, as determined by the near horizon and near boundary expansions of the effective potentials. There are also a few dashed lines plotted whose meaning we will explain in a moment. It is apparent that by going to higher dimensions, a disconnected region close to the apex of the stability region is now causal as well. This means that such a region satisfies all our previous constraints, and yet it will have an arbitrarily low shear viscosity. This is because, as we have seen earlier, the apex of the stability region actually lives on the surface $\eta=0$ determined by the stability constraint in the shear channel. This point lies inside the causality allowed region for $d \leq 10$ \cite{Camanho2010a}.

While in principle there is nothing wrong with having a small ratio, this is not the end of the story. As we have mentioned before, by considering the full effective potentials we can see that they develop causality problems and/or instabilities, but now in the interior of the black hole geometry (see figures \ref{bulkinst}).\!\!\!\!
\FIGURE{\includegraphics[width=0.47\textwidth]{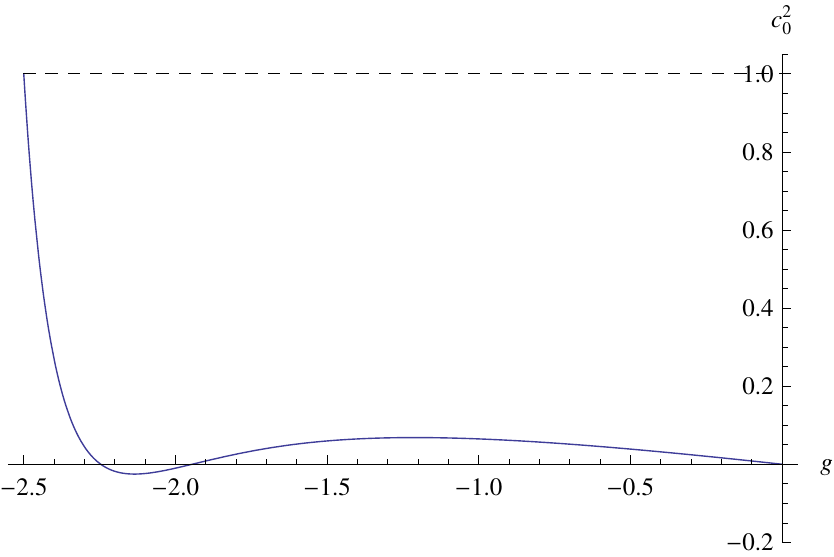}~~\includegraphics[width=0.47\textwidth]{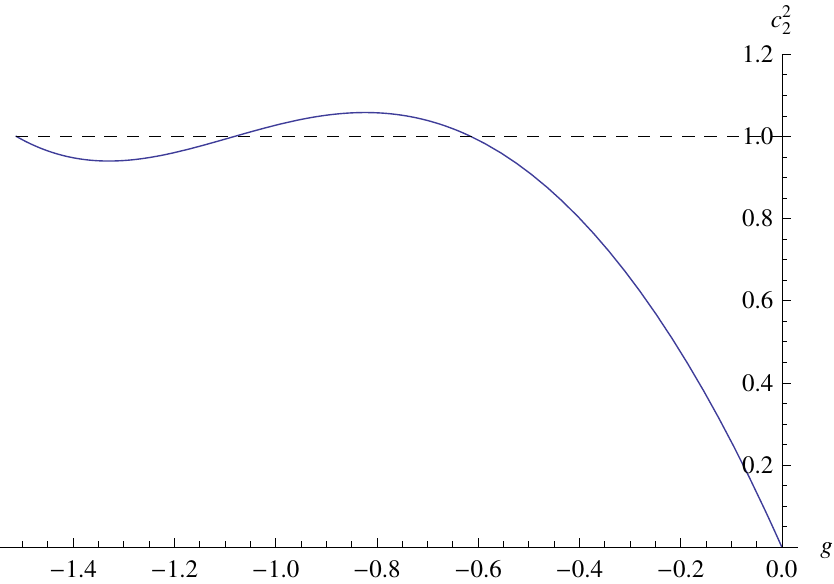}\caption{These figures illustrate the appearance of bulk instabilities (on the left) and bulk causality violation (on the right) for the case of cubic Lovelock theory in ten dimensions. They correspond, respectively, to the sound potential for $\lambda=0.33$ and $\mu=0.108$ and to the tensor potential for $\lambda=0.35$ and $\mu=0.25$. The plot of the potential ends at the boundary where the function $g$ takes the value of the cosmological constant, respectively $\Lambda=-2.5$ and $\Lambda=-1.512$.}
{\label{bulkinst}}}
In the left figure, even though the potential initially rises close to the horizon ($g=0$), it dips into negative values. In the same way, in the figure on the right, the potential goes below one close to the boundary and then makes a hump above this value, leading to causality violation. Such features cannot be fully seen in a perturbative  analysis, although their presence is easily guessed at. Consider for instance moving along the curve where the causality bound is saturated in the tensor channel. This means that the effective potential is of the form
\be
V_{eff} = 1 + 0 \times e^{x} + v_4\, e^{2x} + \ldots
\ee
It is clear then that along this curve we will see causality violation if $v_4$ ever becomes positive. Beyond the point in parameter space where this occurs, there will be a competition between the $e^x$ and $e^{2x}$ terms in the potential expansion which will determine the shape of the causality curve from that point onwards. In general it is clear that to determine this curve one must consider the full effective potential, since all terms in the near-boundary expansion become of the same order. Analogous statements hold for the instability analysis near the horizon. These curves are determined numerically and shown as dashed lines in figures \ref{StableCausal-9d} and \ref{StableCausal-11d}. We can see, in particular, that the sound mode instability curve neatly cuts off the region where $\eta/s$ might become too small, thereby imposing an effective lower bound on this ratio. 

Going to higher order Lovelock theories, one may wonder if this general reasoning will hold. Could the addition of further couplings allow us to escape even the full causality and stability bounds? This does not seem to be so. Evidence for this conclusion comes from an analysis of quartic Lovelock theory, to which we now turn. 

\subsection{Quartic Lovelock theory}
\label{love4}

Quartic theory corresponds to the introduction of an extra non-zero parameter, $a_4 \equiv \frac{\nu}{4}L^6$. The theory exists in $d\geq 9$, and examining it requires solving for quartic polynomials. The analysis of the parameter space of this theory is made in a somewhat similar manner to the cubic case, and we will not go into great detail here. There are resemblances and differences to the cubic theory. For instance, as in the lower order cases the boundary of the excluded region is just the surface where the cosmological constant associated with the relevant branch becomes degenerate. However, this surface ends at one of the triply degenerated lines (three cosmological constants equal) and we need another surface to actually close the excluded region. This also happens in the cubic case where the excluded region is closed by $\mu=\lambda^2$, which is the line along which all three solutions become degenerate at some point inside the geometry.\!\!\!
\FIGURE{\includegraphics[width=0.76\textwidth]{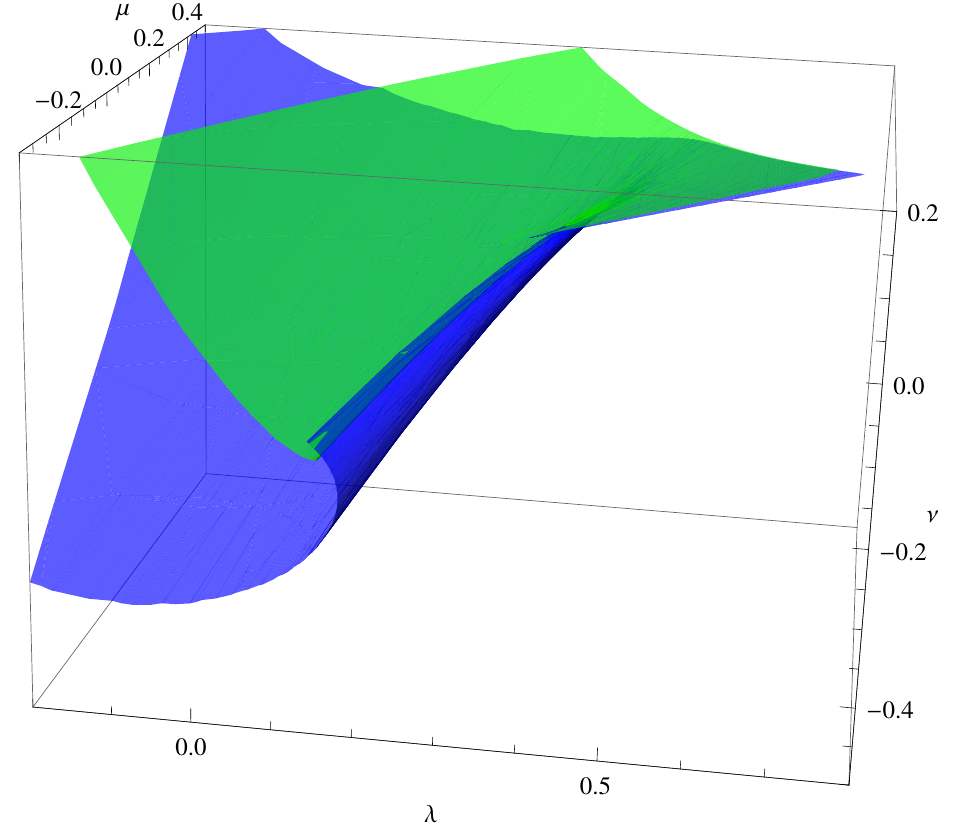}\caption{Surfaces delimiting the (boundary) causal region for the helicity zero and two modes in $d=9$ dimensions; respectively $t_2=-7/5$ (blue) and $t_2=7$ (green). We omit the excluded region boundary surface for the sake of clarity.}
{\label{LL4-causality}}}
In the quartic case the situation is analogous and the excluded region is closed by the surface swept by the line where three solutions degenerate at some value of the radius. This surface cuts off the other surfaces bounding the causal regions for the different helicities: green for tensor channel and dark blue for the sound. As always, the causal region is contained between the tensor and sound surfaces (figure \ref{LL4-causality}). Besides, we also realize that the values of $\lambda$ are not bounded (neither from below nor from above) just due to causality -- by increasing $\nu$ it is possible to go to arbitrarily high $\lambda$.

This situation drastically changes once we consider stability and causality constraints (see figure \ref{LL4-causalstable}).\!\!\!
\FIGURE{\includegraphics[width=0.76\textwidth]{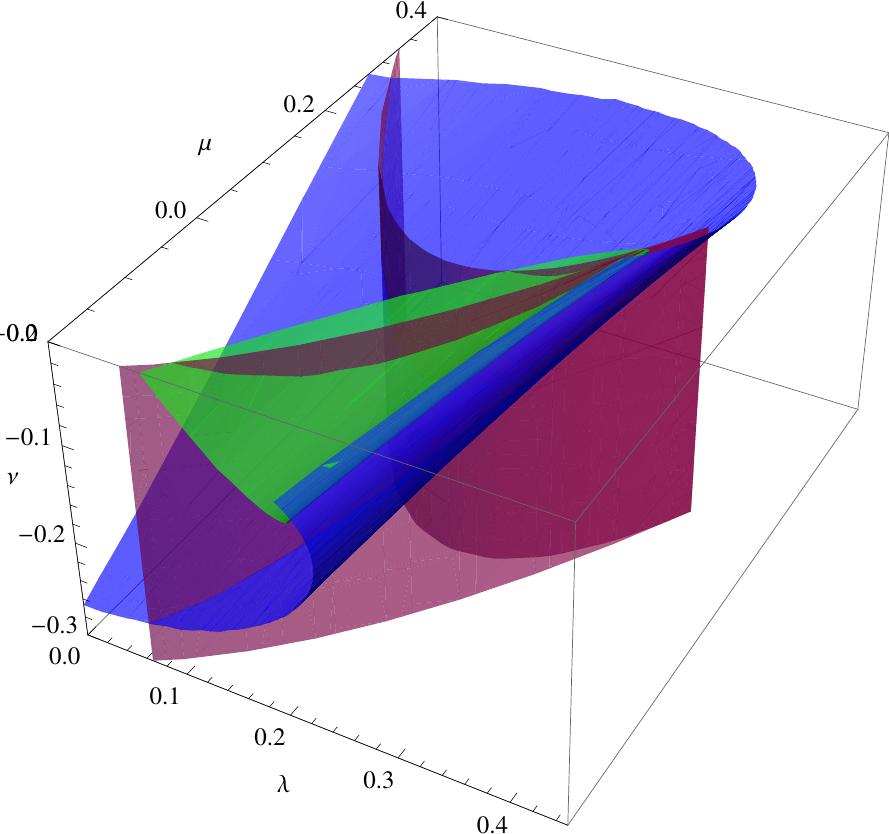}\caption{Stability close to the horizon (in purple) and (boundary) causality surfaces for $d=9$. It can be observed that there is a channel connecting the two disconnected regions found in cubic Lovelock theory (see figure \ref{StableCausal-9d}).}
{\label{LL4-causalstable}}}
When horizon stability is imposed, the allowed region of parameters is further constrained by the purple surfaces; notice, in particular, that it includes a channel connecting the Einstein-Hilbert point with the $\eta/s=0$ line (the apex of the stability region, in purple). We are therefore in a situation analogous to the one in the cubic theory, and we expect the full stability and causality constraints to rescue us. This is indeed the case. In figure \ref{slices9d} we show the causality and stability constraints taken at different $\nu$ slicings. As $\nu$ increases, the causality curves move up in the plot until at some point they block the path to the critical point where $\eta/s=0$. For lower values of $\nu$, instead, we cannot reach this point due to the full sound stability constraint. 
\FIGURE{
\includegraphics[width=0.43\textwidth]{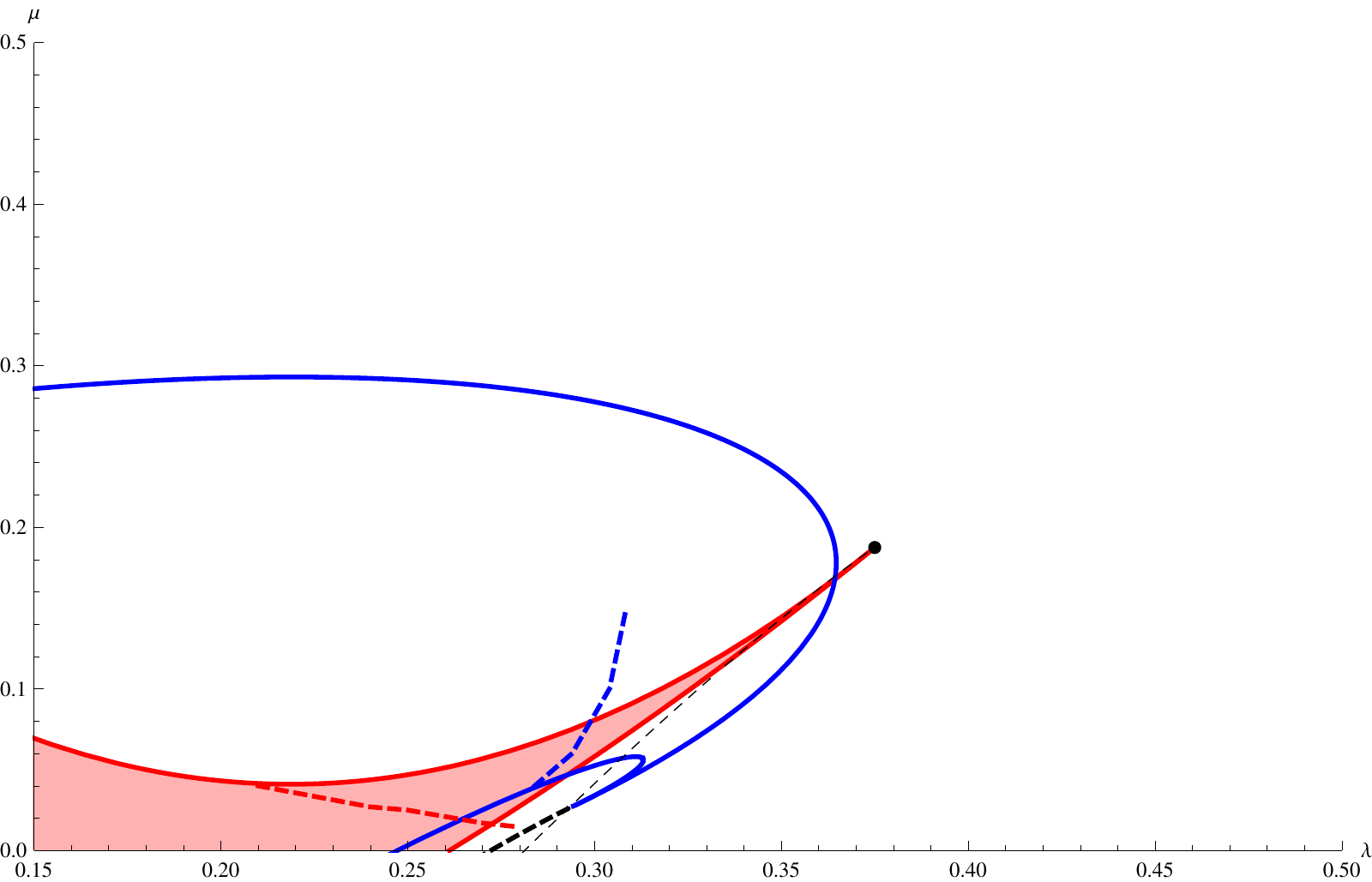} \qquad \includegraphics[width=0.43\textwidth]{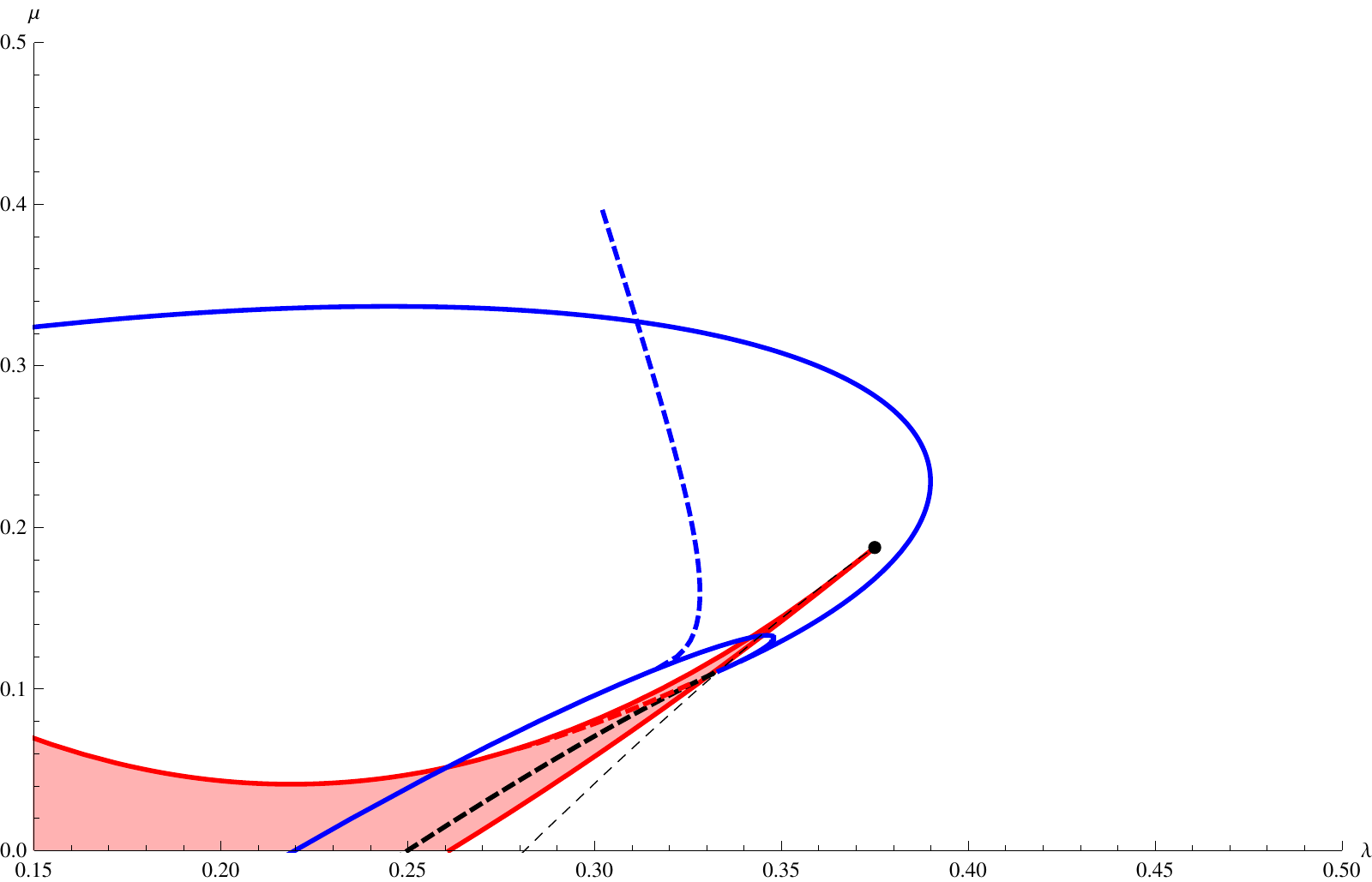} \\
\includegraphics[width=0.43\textwidth]{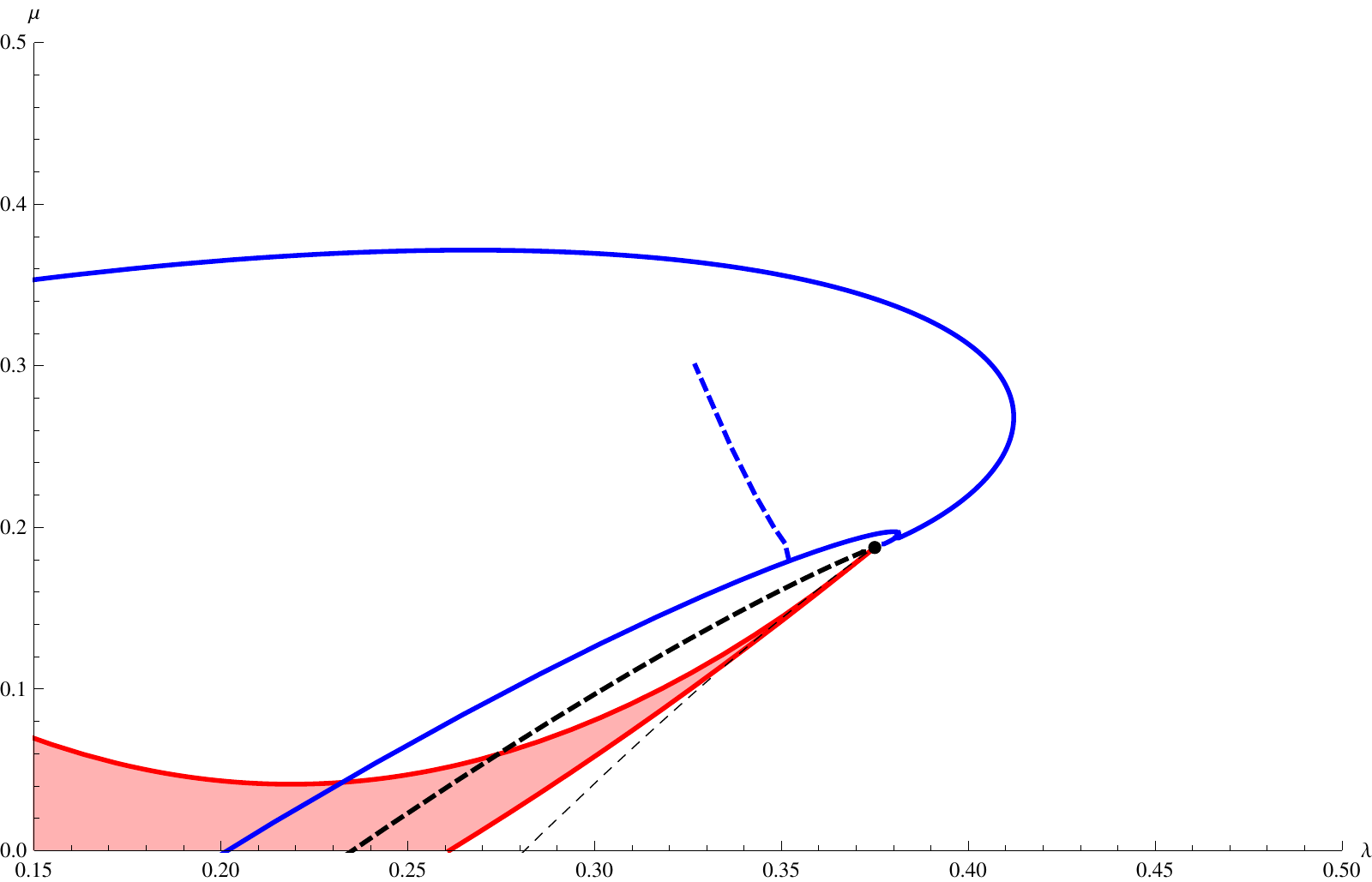}
\caption{Different slices for the quartic case in 9d. Top left corresponds to $\nu=-0.02$, top right $\nu=0$ and below $\nu=1/64$. The black thin dashed line is an imaginary line connecting the different end points of the causality constraints. The thick black line is the boundary of the excluded region. The dashed red and blue lines are the full sound stability and tensor causality curves respectively.}
{\label{slices9d}}}

We have actually done this analysis for various $\nu$ and have reached the same conclusions. These results add further evidence that for every dimension, stability and causality place an effective lower bound on the value of the shear viscosity to entropy ratio.

\subsection{Expansions at $\eta/s=0$}

We would now like to present a general argument that shows one should expect instabilities to appear generically as the $\eta/s=0$ point is approached. The argument is related to the fact that in the section 4, there is something which might invalidate the analysis. This is when the perturbative expansion on the radius, close to the horizon, breaks down. Notice that the tensor and sound mode effective potentials, $c_2^2$ and $c_0^2$ in \reef{potentialsF} have $(d-3)/(d-1)-F'(x)$ in the denominator. At the horizon we have,
\be
\frac{d-3}{d-1} - F'(x) \simeq 2 (\lambda_c-\lambda) ~.
\ee
This means that the near-horizon expansion breaks down whenever $\lambda\simeq \lambda_c$, precisely where $\eta/s\simeq 0$. Close enough to this point the linear analysis breaks down, and we must treat this case separately. This is true for any Lovelock theory in any dimension, and it is therefore of interest to see if there are any general statements one can make. 

Let us then investigate the behavior of the effective potentials given in \reef{pots} at the horizon. First notice that the equation (\ref{eqg}) fixes
\begin{equation}
g(x) \simeq \frac{x}{L^2} + \mathcal O(x^2) ~,
\end{equation}
and we can expand
\begin{equation}
\left( \frac{d-3}{d-1} - F'(x) \right) \approx \left( \frac{d-3}{d-1} - F'(0) \right) - \sum^{}_{n=1} \frac1{n!}\, F^{(n+1)}(0)\; x^{n} ~.
\label{denominator}
\end{equation}
The expansion is such that each term in the above is controlled by a different Lovelock parameter.
For instance $F'(0)=2\lambda$, and so the first term vanishes whenever $\lambda = \lambda_c$. When the following derivatives $F^{(j)}(0)$ vanish, $\forall j<n$, we can actually write
\begin{equation}
F^{(n)}(0) = \Upsilon^{(n+1)}[0] - \prod_{i=1}^{n} \frac{d-2i-1}{d-1} ~,
\label{denomina}
\end{equation}
and so we can set the following term to zero by choosing
\begin{equation}
(a_{n+1})_c = \frac{\Upsilon^{(n)}[0]}{n!} = \frac{1}{n!} \prod_{i=1}^{n-1} \frac{d-2i-1}{d-1} ~.
\end{equation}
In particular,
\begin{equation}
\mu_c =  \frac12 \frac{(d-3)(d-5)}{(d-1)^2} ~, \qquad \nu_c = \frac1{6} \frac{(d-3)(d-5)(d-7)}{(d-1)^3} ~, \qquad \ldots
\end{equation}

For a given (odd) dimensional Lovelock theory, if we set all $K=\frac{d-1}2$ couplings to these particular values, we will be at the so-called AdS Chern-Simons point \cite{Zanelli2005}. Now, notice that (\ref{denominator}) is actually the denominator of the helicity two and helicity zero potentials (and proportional to that of the helicity one mode). The leading correction to those potentials close to the horizon, when we set $F'(0)=2\lambda_c$ and $F^{(j<n)}(0)=0$, is then
\begin{eqnarray}
c_2^2(x) & \approx & - \frac{d-1}{d-4}\; \frac{n-1}{L^2\,\Lambda} + \mathcal O(x) ~, \\ [0.7em]
c_1^2(x) & \approx & - \frac{d-1}{d-3}\; \frac{F^{(n)}(0)}{(n-1)!\,L^2\,\Lambda}\; x^n\; (1 + \mathcal O(x)) ~, \label{horexp2} \\ [0.7em]
c_0^2(x) & \approx & \frac{d-1}{d-2} \; \frac{n-1}{L^2\,\Lambda} + \mathcal O(x) ~.
\end{eqnarray}
The effective potentials have therefore discontinuous limits at the horizon as one approaches $\lambda = \lambda_c$. Since $\Lambda<0$, the sound channel potential tends to a negative constant at the horizon as the critical value $\lambda_c$ is approached\footnote{A caveat to this conclusion is the case where one takes the full function $F'(x)-(d-3)/(d-1)$ to zero, what actually corresponds to the Chern-Simons point. However, it is easy to see from their definition that in this case one of the $c_2^2$ and $c_0^2$ potentials becomes negative.}. We conclude that {\it for any given Lovelock theory, it is impossible to get arbitrarily close to $\eta/s=0$ without running into an instability.} This means that, at least within Lovelock theories of gravity, it seems to be impossible to obtain arbitrarily small values for the $\eta/s$ ratio at any given dimensionality. 

These arguments say nothing about the value of the minimum itself. It seems clear that for any Lovelock theory there will be a minimum, but it is also reasonable to expect that for higher dimensionalities, the existence of extra free parameters would allow a lower value to be reached. For instance, looking naively into the horizon expansion of the effective potential
\be
V_{eff} =\alpha x+\beta x^2+\ldots
\ee
it seems that one could move towards higher values of $\lambda$ staying inside the stability region by simply picking parameters such that $\alpha$, $\beta$, etc. are always negative. In practice, however, we necessarily run into difficulties with this reasoning because of the breakdown of the perturbative expansion as we approach the critical value $\lambda_c$.\footnote{In other words, even though the first term in the horizon expansion might lead to a positive effective potential, the higher order terms might reverse this tendency, and they are becoming more and more important as $\lambda\to \lambda_c$.} In general, the precise interplay between these two issues must be determined by numerics, on a case by case basis.

\section{The $\eta/s$ ratio in higher order Lovelock theories}

\subsection{The minimum ratio in quartic theory}

In Gauss-Bonnet gravity and cubic Lovelock theory, the parameter space is sufficiently small that it can be explored easily. In particular, one can find the full numerical causality and instability curves which determine the allowed regions in parameter space. As one goes to higher orders this analysis becomes more and more baroque, as the hypersurfaces in parameter space delimiting the causal and stable regions are more complicated and cannot be visualized in general. However, as was clear from section \ref{love4} we have succeeded in studying in detail quartic Lovelock theory in various dimensions. For each theory and dimensionality we have found the minimum value for the shear viscosity to entropy ratio, and plotted the results in figure \ref{boundslambda}.\!\!\!
\FIGURE{\includegraphics[width=0.5\textwidth]{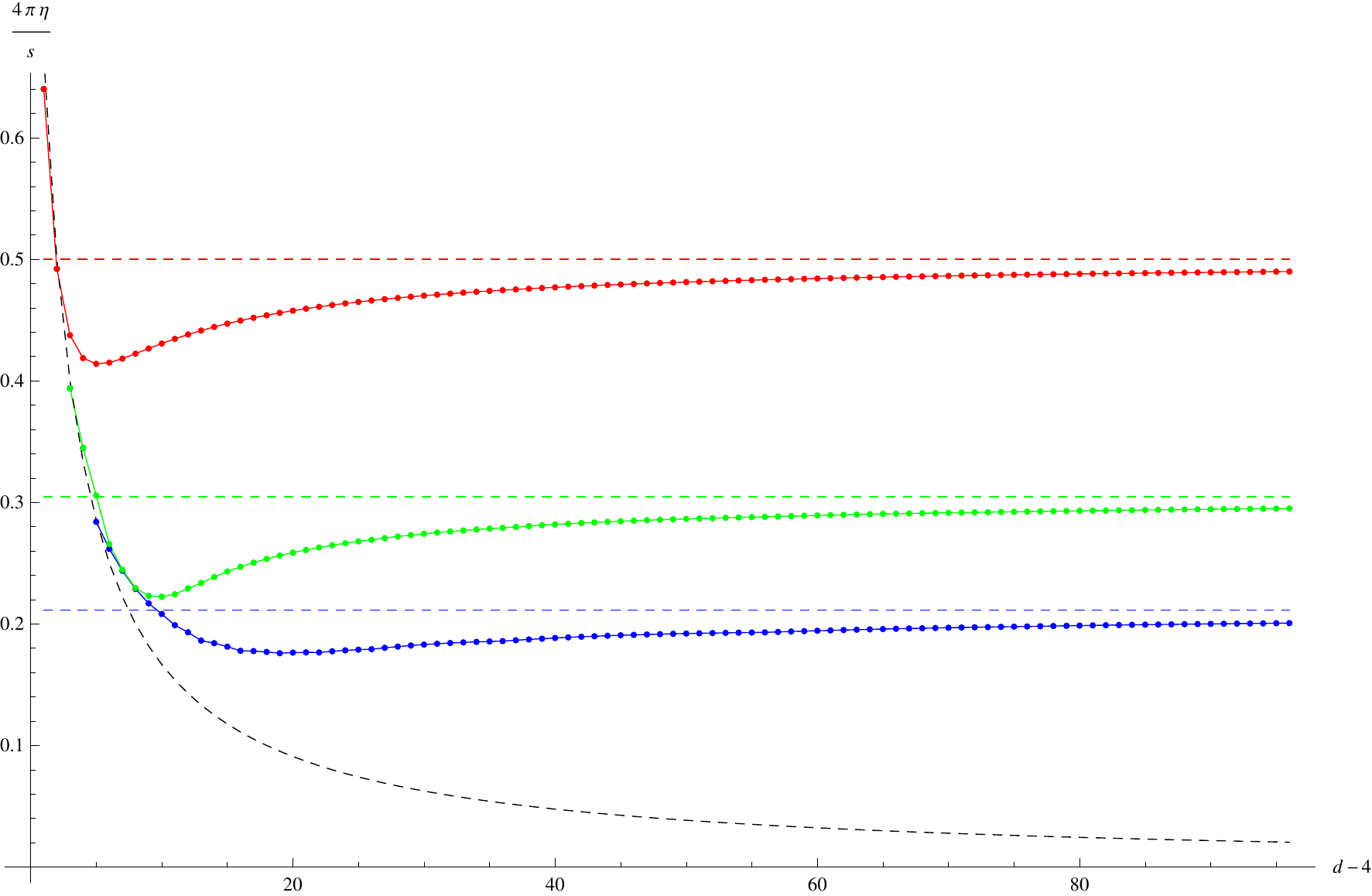}\qquad\includegraphics[width=0.43\textwidth]{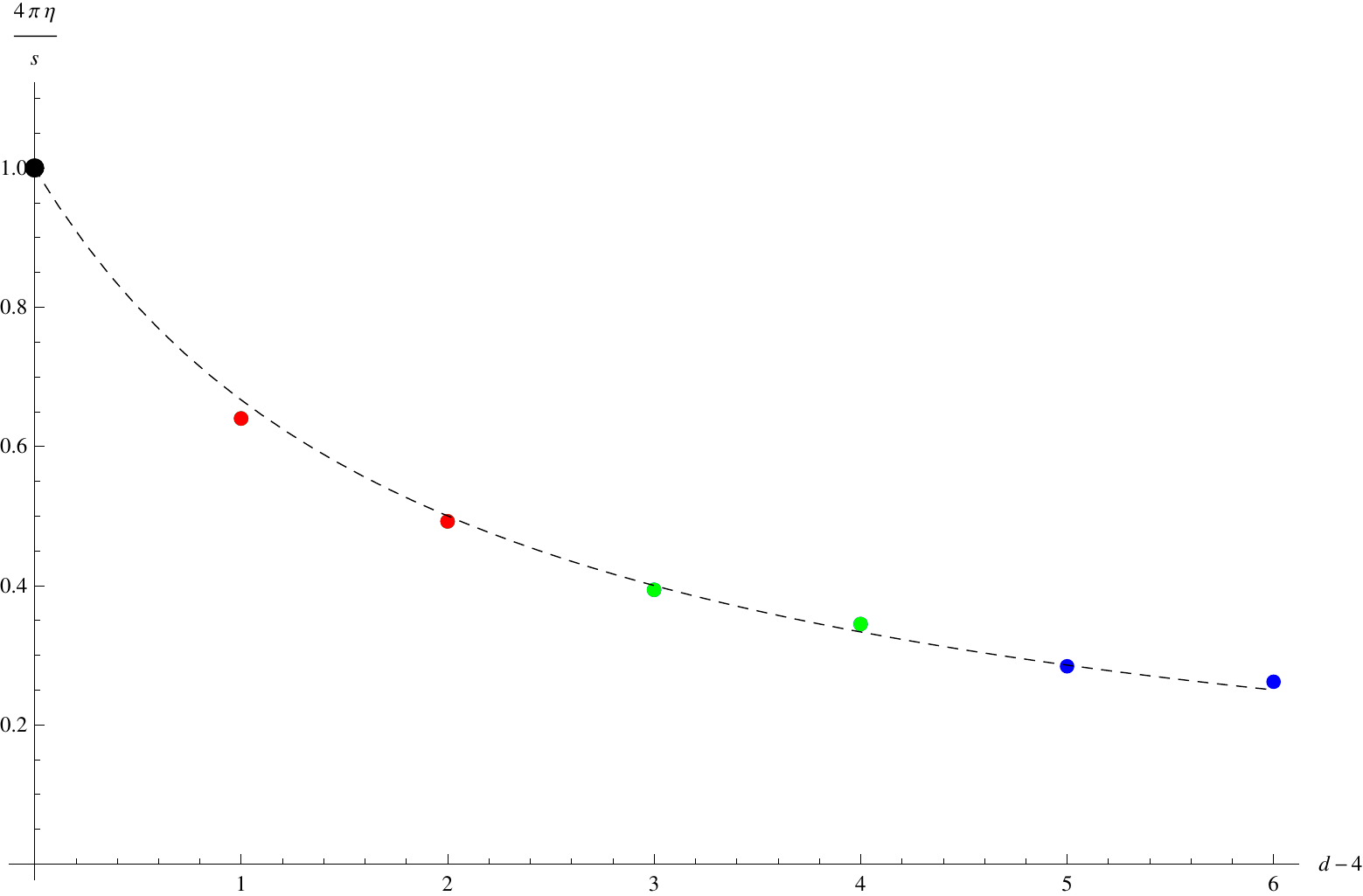}
\caption{ The dots correspond to the numerical (some analytic) bounds for Gauss-Bonnet (red), cubic (green) and quartic (blue) Lovelock theories, and the colored dashed lines correspond to the estimated asymptotic value on each case.  The dashed black line corresponds to the curve $4\pi\eta/s = 2/(d-2)$ that nicely fits (see the zoomed figure at the right) the first seven absolute bounds (including the unit value for $d=4$) and has a tantalizing behavior, $4\pi\eta/s \sim 2/d \sim 1/K_{\rm max}$, for $d \to \infty$, where $K_{\rm max}$ is the integer part of $(d-1)/2$.}
{\label{boundslambda}}}
The first thing we can notice is the remarkable smoothness either for each separated value of $K$ as well as extending from one value $K$ to $K+1$. There is nothing preventing the actual values to jump sharply given the discreteness in the number of couplings involved in these theories. Interestingly, we find that there is a minimum as a function of dimension for any given order. The critical dimension for these minima and the corresponding values of the minimum $\eta/s$ for each theory are quoted in table \ref{tab:a}. 
\begin{table}
	\centering
		\begin{tabular}{c|c|c}
			$K$ & $d_{min}$ & $(4\pi\eta/s)_{min}$ \\
			\hline
			\hline			
			2 & 9 & $219/529\approx 0.414$\\
			3 & 14 & $0.222$\\
			4 & 23 & $0.176$
		\end{tabular}
\caption{Critical dimension ($d_{min}$) for which the minimum value of the viscosity to entropy density ratio is attained for each order in Lovelock ($K$) theory, as well as the corresponding actual minimum value of the quantity $(4\pi\eta/s)_{min}$.}
\label{tab:a}
\end{table}
Overall, the minimum $\eta/s$ ratio seems to be decreasing rapidly as one increases the number of couplings.

It is interesting to comment on a very simple function that nicely fits the minimum value of $\eta/s$ for $d < 11$. It reads
\begin{equation}
\frac{\eta}{s} \simeq \frac{1}{4\pi}\,\frac{2}{d-2} ~.
\label{fitformula}
\end{equation}
It is the simplest curve that smoothly interpolates these points. It also has a nice asymptotic behavior, as we will discuss shortly. However, it is important to stress that this is not a (dimension dependent) bound for $\eta/s$. Indeed, this is already clear in the most relevant case given by $d=5$, where $2/3$ is actually (slightly) greater than $16/25$ (see the right zoomed figure \ref{boundslambda}). The expression in (\ref{fitformula}), though, approximately captures the dependence of a dimensional dependent novel bound for $\eta/s$ arising in Lovelock theories.

\subsection{No dimension-independent $\eta/s$ bound}

In higher order Lovelock theories, the parameter space becomes too large and we must resort to a more limited analysis. We have chosen to concentrate on a particular line through parameter space, parameterized by the single parameter $\Lambda$:
\begin{equation}
\Upsilon[g]=1+g+\lambda g^2 +\ldots=\left(1-\frac{g}{\Lambda}\right)\left(1-\frac{g}{\tilde{\Lambda}}\right)^{K-1} ~.
\label{maxdeg}
\end{equation}
All other parameters can be determined from $\Lambda$ ($\tilde \Lambda$ is fixed by the overall normalization). The curve has the property that it starts at the Einstein-Hilbert point and ends at the maximally degenerate point (MDP) of the theory, at $\Lambda=-K$ where the polynomial has a single degenerate root. A detailed analysis of the stability and causality properties of this curve is performed in appendix \reef{appendixB}. We have found that for generic values of $K$ and $d$ there is an interval of $\Lambda$ values where there is instability at the horizon\footnote{Except for $K<7$, where it disappears for high enough $d$. Whenever the unstable interval is not present the whole curve is stable.}. This is the range of values where our curve goes outside the stability wedge. Beyond this interval, our curve returns to the stability wedge, but nevertheless there are still instabilities -- no longer at the horizon but in the bulk. These instabilities persist all the way to the end of the curve, where it reaches the maximally degenerate point.

From these results it is clear that generically, the maximum stable and causal point one may reach along this curve is the first crossing point with the stability wedge. This crossing point can be found analytically, though the results are not particularly enlightening. However, there is a considerable simplification at very large dimensionality. To first approximation the crossing point is then determined by solving the algebraic equation,
\be
K^2 (3+2 \Lambda)+2 K \Lambda(4+3\Lambda)+\Lambda^2(6+6 \Lambda+\Lambda^2)=0 ~.
\ee
Solving this equation, and using \reef{lambdamu}, we can find the value of $\lambda$ at the crossing point. However, this value actually decreases as $K$ increases. This leads to a minimum $\eta/s$ ratio which increases as we include more couplings. Notice there is no contradiction here: recall we are moving along a particular curve in parameter space, which is simply not the optimal one in terms of finding the minimum possible value for this ratio.

Although the curve we have chosen is not optimal in this sense, it is still useful as we can use it as a base point for exploring nearby regions of parameter space. In particular, consider moving along the curve towards the maximally degenerate point. For high enough $d$, it is shown in Appendix \ref{appendixB} that close to this point there is only an instability in the sound channel, i.e. all other channels are causal and stable\footnote{The other relevant constraint is actually the tensor causality one but the causality violating region can be made as small as one wishes just by increasing the dimensionality.}. We now perturb our curve by considering a small perturbation $\delta \lambda\, g^2$ or $\delta \mu\, g^3$ to \reef{maxdeg}.
The resulting potentials can then be easily found numerically.

\FIGURE{\includegraphics[width=0.45\textwidth]{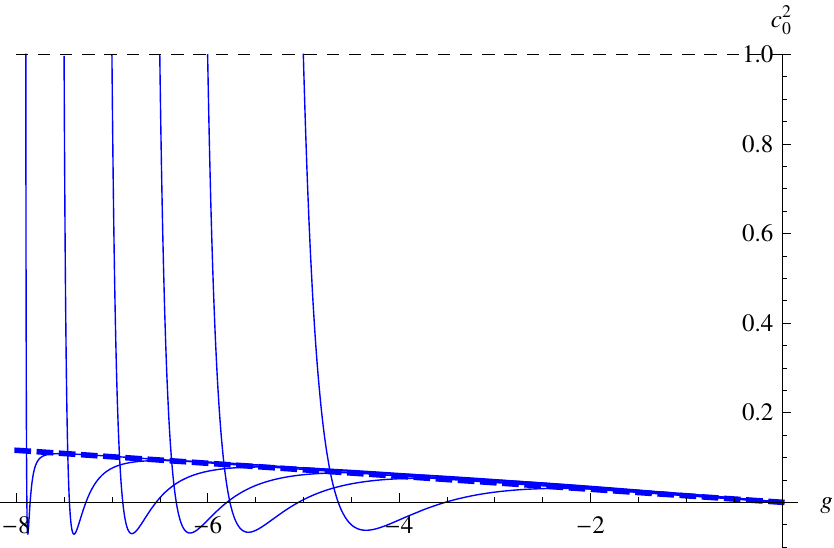}~~\includegraphics[width=0.45\textwidth]{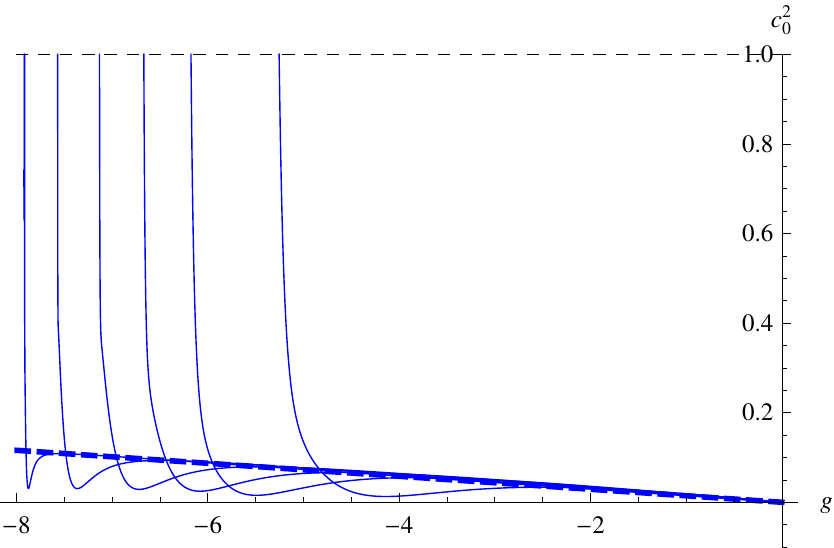}
\caption{On the left, the helicity zero potential for $K=8$, $d=100$ and $\Lambda=-8$ (dashed), $-7.9,-7.5,-7,-6.5,-6,-5$ along our curve. On the right the same curve slightly perturbed by a $\delta \lambda g^2$ term in the polynomial. The precise values of $\delta \lambda$ are $\delta \lambda=0, 1.8\cdot 10^{-18}, 6\cdot 10^{-13}, 1.9\cdot 10^{-10}, 6\cdot 10^{-9}, 8\cdot 10^{-8}, 3\cdot 10^{-6}$.}
{\label{stab0CSpointM}}}

In figure \ref{stab0CSpointM} we show the spin-0 effective potential for $K=8$, for various $\Lambda$ before and after adding a small fluctuation. The dashed curve represents the limiting curve for $K=8$ along our particular trajectory in parameter space ($\Lambda=-8$). On the left-hand figures, we see how although at the MDP itself the potential does not present any instabilities, it is unstable all the way up to that point along our curve, i.e. there are negative potential wells in the bulk. On the right-hand side figure we see how adding just a small perturbation is capable of lifting the instability. This sensitiveness of the potential to fluctuations is due to the fact that the maximally degenerate point is special. At this point we have $\Upsilon^{(n)}(\Lambda)=0$, with $n=0,\ldots, K-1$, which means that the perturbative boundary expansion of the potentials breaks down. Accordingly there is a discontinuous limit at that point. Indeed, taking the $\Lambda\to -K$ limit along our original curve we find that the effective potentials become (assuming $d\neq 2K+1$):
\begin{eqnarray}
c_2^2 & = &-\frac{d-3K-1}{K^2(d-4)} g(x) ~, \nonumber\\[0.9em]
c_1^2 & = &-\frac{d-2K-1}{K^2(d-3)} g(x) ~, \\[0.9em]
c_0^2 & = &-\frac{d-K-1}{K^2(d-2)}  g(x) ~. \nonumber
\end{eqnarray}
Since $g(x)\in [\Lambda,0]$ the above are actually always positive for sufficiently large dimensionality. In particular, although the $c_0$ potential always has a negative region for  $\Lambda$ larger than but close to $-K$, it is actually tending discontinuously to a perfectly reasonable potential. This is clear from figure \ref{stab0CSpointM}.

Our results then indicate that it is reasonable to expect that sufficiently close to the MDP there should be a trajectory in parameter space which is both stable and causal. We have shown this is certainly true up to $K=8$. If this result holds for higher $K$, then it is always possible to reach sufficiently close to the MDP and have well behaved effective potentials. As such, the theory seems to be well defined at that point.

At the MDP we have
\begin{equation}
\lambda_{\mbox{\tiny MDP}}=\frac{K-1}{2K}
\end{equation}
and therefore a value of the viscosity to entropy ratio
\begin{equation}
\left(\frac{\eta}{s}\right)_{\mbox{\tiny MDP}}=\frac{1}{4\pi}\frac{1}{K}~.
\end{equation}
By taking high enough $K$, this can be made arbitrarily small. Besides, since the maximum $K$, $K_{\rm max}$, is proportional to $d/2$ in the large $d$ limit, we obtain the asymptotic behavior already suggested by the formula (\ref{fitformula}) (see also figure \ref{boundslambda}).

To summarize, we have given evidence that it is possible to find a point in parameter space, sufficiently close to the MDP of a particular Lovelock theory, such that no stability or causality issues occur for high enough $d$. In particular we have checked that it is possible to do this up to $K=8$. We conjecture that this can always be achieved for any $K$, and we therefore come to the conclusion that {\em the viscosity to entropy ratio can be made arbitrarily small by considering a Lovelock theory of high enough order}. In other words, there is no bound for the $\eta/s$ ratio which is independent of the dimensionality, at least in the class of Lovelock theories of gravity. This goes in line with our expectations that adding more and more couplings, or free parameters in the Lagrangian, it should be possible to reach lower and lower values for the shear viscosity to entropy ratio. 

\section{Discussion}

In this paper we have considered various aspects of arbitrary Lovelock theories of gravity, both at zero and finite temperature. For these theories, we have explicitly computed the parameters $t_2$ and $t_4$ appearing in the Maldacena-Hofman parameterization of the $3$-point function of the stress tensor \cite{Hofman2008}, and we have checked that they satisfy the positivity of energy bounds conjectured in that paper. These results constitute a proof of previous conjectures in the literature \cite{Boer2010,Camanho2010a}.

It is intriguing that classical Lovelock gravity in AdS space can capture highly non-trivial aspects of conformal field theory physics. While Einstein gravity is known to describe the universal spin-2 sector of a large class of strongly coupled CFTs, no dual to Lovelock gravity (beyond GB) has been found, nor have the Lovelock terms with finite coefficients appeared in string theory: While higher powers of curvature generically appear in string theory as $\alpha'$ corrections to the supergravity action, these corrections are necessarily perturbative. In Lovelock theory we have dimensionful parameters appearing in the lagrangian, which do not appear to be protected from quantum corrections by any symmetry. That from the classical theory we can obtain specific numerical predictions on which values of the parameters should or not be allowed by causality and that these restrictions should agree with analogous restrictions in a hypothetical dual CFTs is highly suggestive. We obtain what seem to be splendid checks of the AdS/CFT correspondence, the results being smooth functions of the spacetime dimensionality.

One possible hint is that Lovelock theories all satisfy $t_4=0$, which is required for supersymmetric CFTs \cite{Kulaxizi2010}. For free field theories, $t_4=0$ is obtained by picking the appropriate supersymmetric matter content. In Lovelock gravity, the $t_2$ and $t_4$ parameters can be obtained in the boundary expansions of the effective potentials \reef{pots}. The $t_4=0$ result should imply some relationship between these. Indeed, it is easy to see that from \reef{pots} we have
\be
(d-4)\, c_2^2(r) + (d-2)\, c_0^2(r) - 2 (d-3)\, c_1^2(r) = 0 ~,
\label{holsusy}
\ee
which is actually valid at any radius $r$ and not just at the boundary, where it indeed implies $t_4=0$.  Beyond the boundary expansion, the relationship written above should impose constraints on higher $n$-point functions of the theory. It is tempting to conjecture that these are the holographic duals of some sort of supersymmetric constraint.

An important part of our analysis was the study of these effective potentials, searching for causality violation or instabilities in the large momentum limit which might rule out regions of the parameter space. Notice, however, that while the former condition is a {\it bona-fide} constraint on the theory, since it would lead to causality, the second seems to hold only for the validity of the black brane solution, and is not a fundamental restriction. We have found that causality violation can occur deep inside the bulk, and therefore this effect cannot be captured by a boundary expansion. This means that, contrary to previous expectations, causality violation is not necessarily a UV feature of the CFTs related to positivity of energy restrictions. Holographically this means that there are interesting constraints on the parameters of the CFT which cannot be seen in perturbation theory. The field-theoretic origin of such constraints remains a mystery.\footnote{It would be of definite interest to study whether these constraints are related or not to unitarity restrictions that were very recently shown to arise in the computation of $2$-point functions of the stress energy tensor at finite 
temperature and large energy and momenta \cite{Kulaxizi2010a}.} On the gravity side, these restrictions were essential to prevent an arbitrarily small $\eta/s$ ratio for a fixed theory, as seen for the concrete examples of cubic and quartic Lovelock theories.

For higher order theories, we have resorted to the study of a particular curve through parameter space. We chose this curve on the basis of simplicity and the capability of analytic treatment. The upshot of our analysis is that it seems likely that there is some curve leading arbitrarily close to the maximally degenerate point of the theory. The reader might be suspicious about this -- after all, at this point the central charge of the dual conformal field theory is vanishing, and so is the kinetic term of the graviton fluctuations around the AdS black hole. That is why we emphasize that one is really considering a curve leading to this point, but causality constraints keep us at a safe distance where everything is well defined. 

In all cases we considered in detail, the minimum value for the ratio was obtained well away from the maximally degenerate point. In any case, we hope our work stimulates study of the properties of the theory around this particular point. In the case $d = 2 K + 1$, these theories display symmetry enhancement becoming gauge theories of the Chern-Simons group: they have local AdS symmetry whereas all other theories with $d > 2 K + 1$ have local Lorenz invariance \cite{Zanelli2005}. Contrary to what is suggested by the fact that $C_T$ vanishes in the dual conformal field theory, there are some hints pointing towards the existence of interesting physics on these theories. For instance, the thermodynamics of their black holes displays a qualitative difference to that of generic AdS Lovelock black holes: the temperature grows linearly with the horizon radius, and the specific heat is a continuous, monotonically increasing and positive function of $r_+$ \cite{Crisostomo2000}. Thus, Chern-Simons black holes can reach thermal equilibrium with a heat bath at any temperature and they are stable under thermal fluctuations. There is a mass gap between the massless black hole and AdS spacetime, as it happens in the case of $d=3$ \cite{Banados1992a}.

Our results indicate that, in the case of very large $d$, it should be possible to reach a value of the $\eta/s$ ratio of {\em at least}
\be
\frac{\eta}s=\frac{1}{4\pi} \frac{1}{K}
\ee
in the $K$th order theory. Intuitively, this is simply telling us that the more parameters we have, the lower the ratio can become. This is reminiscent of the species counter-argument to a minimum bound on $\eta/s$ \cite{Cohen2007a,Cherman2008}. Of course, in Lovelock gravity one needs to increase the number of dimensions in order to have more parameters. Then it would seem that for all practical purposes there is a bound on $\eta/s$ for any finite dimensionality. This is certainly a possibility, and it is definitely true for Lovelock theory -- a combination of stability and causality insures it. However, even for fixed dimensionality it seems like it might be possible to have interesting higher curvature corrections with finite coefficients \cite{Myers2010c,Oliva2010,Oliva2010a}. These ``quasi-topological gravities'' have exact black hole solutions of a similar nature to those of Lovelock theory, including their thermodynamic features. However, hydrodynamic as well as causality and stability properties, are quite different \cite{Myers2010d}. While there is much work to be done understanding these theories, it seems plausible that a sufficiently high order quasi-topological theory will have a very small $\eta/s$ ratio. 

The stability wedge in all Lovelock theories has a non-trivial shape in the $\lambda$-$\mu$ plane but is otherwise a cylinder in the remaining couplings. This means, in particular, that there is a lower bound for the cubic Lovelock parameter $\mu$. It results from instabilities of the black brane solution and, as such, should be related to properties of the CFT plasma rather than of the zero temperature CFT. The precise meaning of this bound in the field theory side and its consequences in the computation of physical quantities such as transport coefficients or thermodynamic variables, as well as the absence of a higher bound for this parameter, are interesting avenues for further research.

As discussed earlier in the case of GB \cite{Camanho2010} and cubic Lovelock theory \cite{Camanho2010a}, it is tempting to say something about the existence of a lower bound on $\lambda$ (see figure \ref{boundslambda-low}), that implies an upper bound for $\eta/s$. It is due to the causality constraint in the former and the stability constraint on the latter. This seems to be in line with the expectation that the shear viscosity of a strongly coupled system cannot be too large. On this respect, it is interesting to mention that this bound disappears in Lovelock theories of quartic or higher order -- {\it i.e.}, for $d \geq 9$ (see figure \ref{boundslambda-low}) --, since no expected problems as causality violation or instabilities of the sort discussed in this paper arise in that direction. This might be hinting towards the existence of pathologies of a different nature that we have been unable to characterize.
\FIGURE{\includegraphics[width=0.67\textwidth]{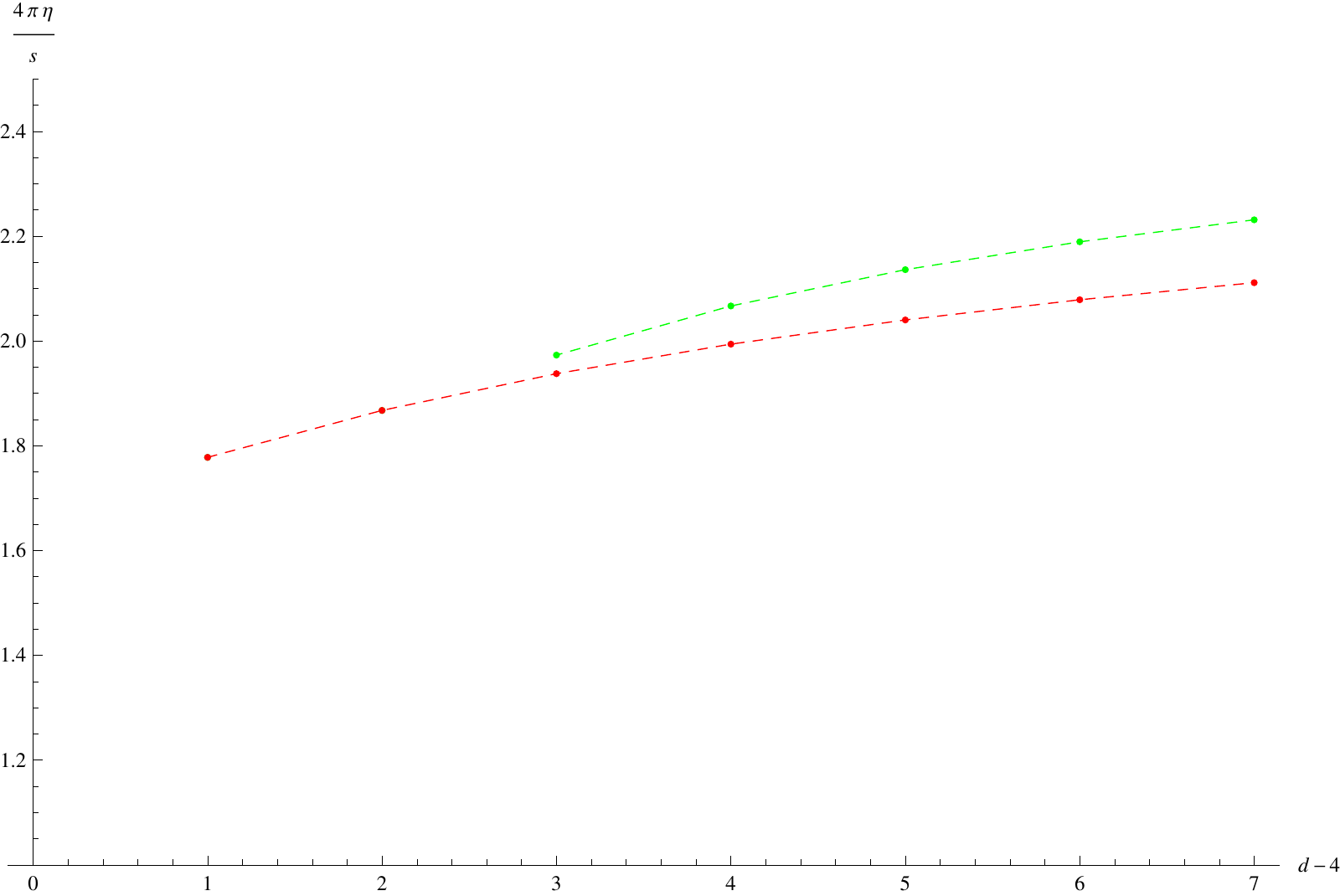}
\caption{ The dots correspond to the numerical (some analytic) $\eta/s$ `upper bounds' for $K=2,3,4$ (red, green and blue respectively).}
{\label{boundslambda-low}}}

To conclude, we hope our work stimulates further research into holographic studies of the $\eta/s$ ratio in particular, and the dynamics of Lovelock theories of gravity and holography in general. It might be useful to explore further the effective potentials. These relatively simple functions seem to encode a great deal of information about the hypothetic dual CFT, information which we are only beginning to extract.

\section*{Acknowledgements}

We have benefitted from most valuable conversations with Diego Hofman, Robert Myers and Aninda Sinha. We would like to thank Rob Myers for a critical reading of the manuscript and interesting comments about it. JDE would also like to thank Prem Kumar, Sameer Murthy, Carlos N\'u\~nez and Yaron Oz for raising interesting issues when a preliminary version of this work was presented. 
MFP acknowledges financial support from the Portuguese Government FCT grant SFRH/BD/23438/2005, from DAMTP, University of Cambridge, and LPTHE, Universit\'e Pierre et Marie Curie, Paris. XOC and MFP also thank Perimeter Institute for their kind hospitality where part of this research was performed. XOC is supported by a spanish FPU fellowship.

This work is supported in part by MICINN and FEDER (grant FPA2008-01838), by Xunta de Galicia (Conseller\'{\i}a de Educaci\'on and grant PGIDIT10PXIB206075PR), and by the Spanish Consolider-Ingenio 2010 Programme CPAN (CSD2007-00042). 
The Centro de Estudios Cient\'{\i}ficos (CECS) is funded by the Chilean Government through the Millennium Science Initiative and the Centers of Excellence Base Financing Program of Conicyt, and by the Conicyt grant ``Southern Theoretical Physics Laboratory'' ACT-91. CECS is also supported by a group of private companies which at present includes Antofagasta Minerals, Arauco, Empresas CMPC, Indura, Naviera Ultragas and Telef\'onica del Sur.

\appendix

\section{Three-point function parameters}
\label{3point}

Conformal symmetry is powerful enough to determine the form of the $3$-point function for the stress-energy tensor up to five constants and this are further constrained by conservation laws allowing us to reduce the number of independent parameters to three, $\mathcal{A}$, $\mathcal{B}$ and $\mathcal{C}$. Further one finds that Ward identities relate the two- and three-point functions and so $C_T$ can be expressed in terms of these three constants.
\begin{equation}
C_T=\frac{\Omega_{d-2}}{2}\frac{(d-2)(d+1)\mathcal{A}-2\mathcal{B}-4d\mathcal{C}}{(d-1)(d+1)}
\end{equation}
where $\Omega_{d-2}$ is the area of the unit $(d-2)$-sphere. Also as the energy flux one-point function is just a quocient of a three- and a two-point function $t_2$ and $t_4$ should also be writable in terms of the three parameters $\mathcal{A}$, $\mathcal{B}$ and $\mathcal{C}$. This was done in \cite{Buchel2010a} yielding
\begin{eqnarray}
t_2&=&\frac{2d}{d-1}\frac{(d-3)(d+1)d\mathcal{A}+3(d-1)^2\mathcal{B}-4(d-1)(2d-1)\mathcal{C}}{(d-2)(d+1)\mathcal{A}-2\mathcal{B}-4d\mathcal{C}} ~, \nonumber \\ [0.9em]
t_4&=&-\frac{d}{d-1}\frac{(d+1)(2(d-1)^2-3(d-1)-3)\mathcal{A}+2(d-1)^2(d+1)\mathcal{B}-4(d-1)d(d+1)\mathcal{C}}{(d-2)(d+1)\mathcal{A}-2\mathcal{B}-4d\mathcal{C}}~,\nonumber
\end{eqnarray}
Then the scalar, vector and tensor constraints coming from from the positivity of energy can also be translated into
\begin{eqnarray}
-\frac{d}{d-1}\frac{(d-3)(d+1)\mathcal{A}+2(d-1)\mathcal{B}-4(d-1)\mathcal{C}}{(d-2)(d+1)\mathcal{A}-2\mathcal{B}-4d\mathcal{C}}&\geq& 0 ~,\nonumber\\[0.9em]
d\frac{(d-3)(d+1)\mathcal{A}+(3d-5)\mathcal{B}-8(d-1)\mathcal{C}}{(d-2)(d+1)\mathcal{A}-2\mathcal{B}-4d\mathcal{C}}&\geq& 0 ~,\\[0.9em]
-2d^2(d-2)\frac{\mathcal{B}-2\mathcal{C}}{(d-2)(d+1)\mathcal{A}-2\mathcal{B}-4d\mathcal{C}}&\geq& 0 ~.\nonumber
\end{eqnarray} 
constraints that can be identified with those coming from causality in the boundary theory in order to obtain expressions for $\mathcal A, \mathcal B,\mathcal C$. However in order to unambiguously determine the normalization of these coefficients we can just obtain them in terms of $C_T, t_2, t_4$ and plug the expressions for these into them. We obtain then
\begin{eqnarray}
\mathcal{A} &=& - \frac{(d-1)^3}{(d-2)^3}\, \frac{\Gamma[d]}{\pi^{d-1}}\, \frac{1}{(-\Lambda)^{d/2}} \left(\frac{2  d \Lambda  \Upsilon''[\Lambda]}{(d-4)^2}+ \Upsilon'[\Lambda]\right) ~, \nonumber \\[0.9em]
\mathcal{B} &=& -\frac{  (d-1) }{(d-2)^3 }\frac{\Gamma[d]}{\pi^{d-1} } \frac{1}{(-\Lambda)^{d/2}}\left(\frac{ (d-1) d \left(d^2-4 d + 6\right) \Lambda  \Upsilon''[\Lambda]}{(d-4)^2}+ \left(d^3-4
d^2+5 d-1\right) \Upsilon'[\Lambda]\right) ~,\nonumber\\[0.9em]
\mathcal{C} &=& -\frac{(d-1)^2 }{2(d-2)^3}\frac{\Gamma[d]}{\pi^{d-1}} \frac{1}{(-\Lambda)^{d/2}}\left(\frac{ \left(d^3-3 d^2+3 d-4\right) \Lambda  \Upsilon''[\Lambda]}{(d-4)^2}+\frac{1}{2}  (2 (d-3) d+3) \Upsilon'[\Lambda]\right) ~. \nonumber
\end{eqnarray}
%

\section{A curve through the parameter space of Lovelock gravities}
\label{appendixB}

Exploring the full parameter space for a generic Lovelock theory is fairly untractable. As such, we choose to consider a particular curve through parameter space, simple enough that we can determine where along it a given theory preserves causality and stability. This curve corresponds to such a choice of parameters that the defining polynomial of the $K$-th order theory becomes (we set $L=1$ here for simplicity)
\begin{equation}
\Upsilon[g]=1+g+\lambda g^2 +\ldots=\left(1-\frac{g}{\Lambda}\right)\left(1-\frac{g}{\tilde{\Lambda}}\right)^{K-1} ~,
\label{maxdegb}
\end{equation}
where there is just one free parameter as
\begin{equation}
\tilde{\Lambda}=-\frac{(K-1)\Lambda}{1+\Lambda} ~,
\label{cosmolambda}
\end{equation}
to ensure that the coefficient of order $g$ on the polynomial (the Einstein-Hilbert term) is actually one. All the Lovelock coefficients can be written in terms of this parameter,
\begin{eqnarray}
\lambda&=&\frac{(1+\Lambda)((K-2)\Lambda-K)}{2(K-1)\Lambda^2} ~, \nonumber \\[0.9em]
\mu&=&\frac{(1+\Lambda)^2(K-2)((K-3)\Lambda-2K)}{2(K-1)^2\Lambda^3} ~, \qquad \ldots
\label{lambdamu}
\end{eqnarray}
We call this from now on the {\sl maximally degenerated trajectory} (MDT). In figure \ref{trajectory} we show the projection of our curve on the $\lambda$-$\mu$-plane for the particular case $K=5$. For the range $\Lambda\in[-K,0]$, the cosmological constant, corresponding to the root with minimum absolute value of $\Upsilon=0$, is given by $\Lambda$, and beyond that by $\tilde \Lambda$. The curve parameterized by $\Lambda$ connects Einstein-Hilbert gravity ($\Lambda=-1$) with the maximally degenerated Lovelock theory ($\Lambda=\tilde{\Lambda}=-K$), henceforth denoted by the maximally degenerate point (MDP). 
\FIGURE{\includegraphics[width=0.8\textwidth]{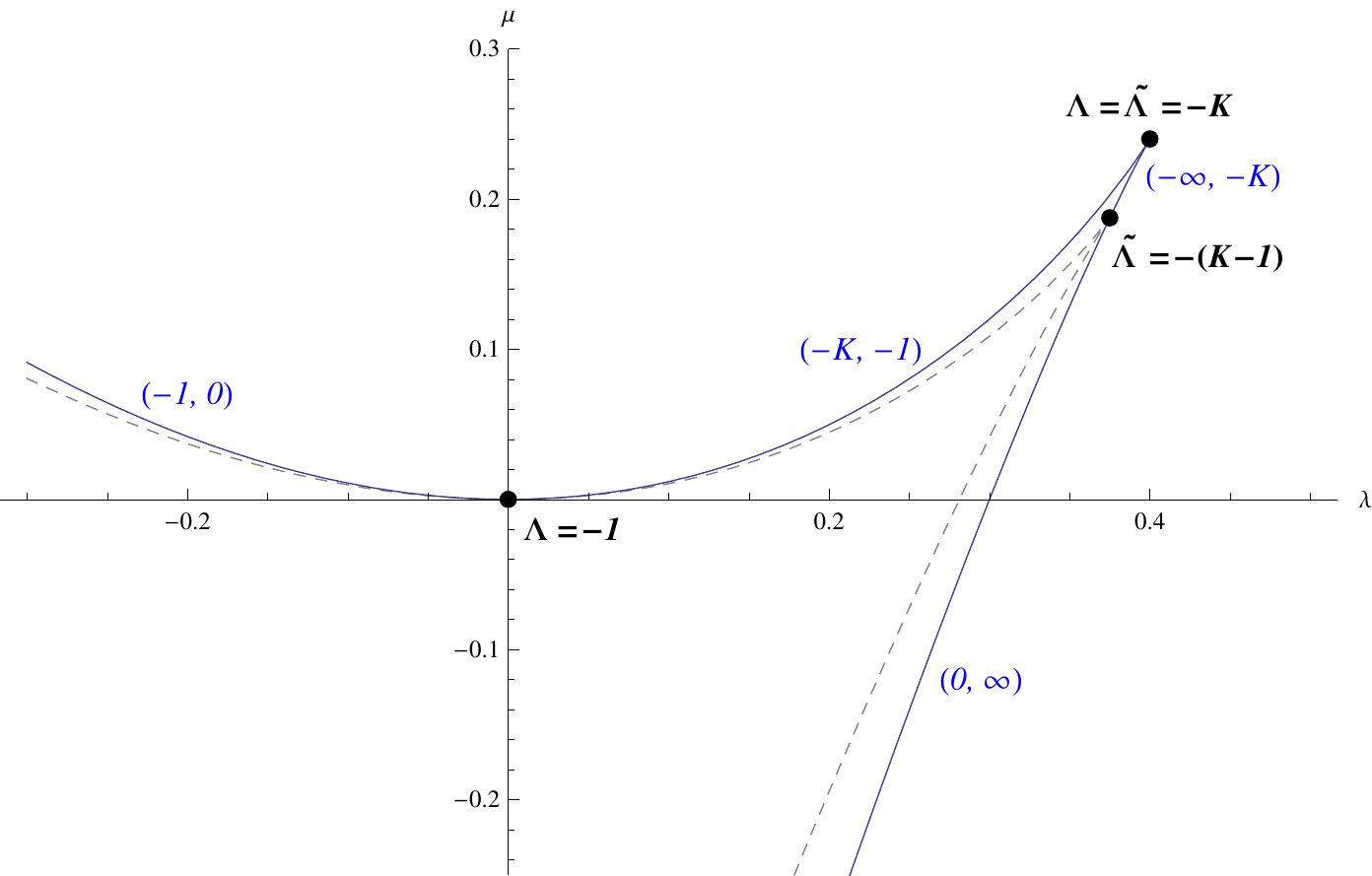}\caption{Projection of the points corresponding to (\ref{maxdegb}) in the $\lambda$-$\mu$-plane for $K=5$. The dashed line correspond to the projection of the $(K-1)$th order trajectory, and in blue the corresponding interval of $\Lambda$ values for each part of the curve is indicated.}
{\label{trajectory}}}
Given this curve, we want to see where instabilities or causality violation may occur. We will restrict ourselves to the range $\Lambda\in(-K,-1)$ section for any value $K$, {\it i.e.}, to the part of the trajectory connecting the Einstein-Hilbert point with the MDP.
We are looking for solutions of $c_i^2(r)$ equals zero or one for the stability or the causality analysis respectively. These constraints can be written as polynomial equations in $g$, and the appearance of new roots is signaled by a change of sign of the corresponding discriminant. An important point worth noticing here is that we are just interested in roots, $g_{\star}$, in the interval of allowed values for the function $g$ 
\begin{equation}
g_{\star}\in[\Lambda,0] ~,
\label{grange}
\end{equation}
where $\Lambda$ is the closest to zero root of $\Upsilon[\Lambda]=0$. The zeros of the discriminant separate the theory space spanned by $(d,K)$ into distinct regions. By examining the properties of curve for a particular pair $(d,K)$ we are guaranteed to have found the generic behaviour of the curve within the region containing that point.

\subsection{Stability analysis}

All three potentials are bound to yield zero at the horizon and one at the boundary. Thus, any root of $c_i^2[g]=0$ coming from outside the interval \reef{grange} has to enter it through the horizon ($g=0$) and in that case the horizon (which is always a zero for the equation) becomes a degenerate root. This is exactly what we studied when analyzing stability at the horizon -- we look for the vanishing of the first coefficient in the horizon expansion. The same phenomenon occurs at the boundary for the causality analysis. Conversely, if any root leaves the interval between the boundary and the horizon one will also see a degenerate root appearing. The only other option for an instability to show up is as a new double root appearing in the bulk. In any case, it is clear that  a study of the discriminants of the polynomial equations tells us exactly when these double roots occur. By carefully studying the regions in parameter space where the discriminant changes sign, one can learn where the stable and/or causal regions lie in parameter space. While this program is very involved in the generic case, the particular trajectory in parameter space we have chosen is simple enough so that this analysis of stability and causality can be done for all three potentials.

\subsubsection{Stability at the horizon}

The horizon stability analysis can be performed by combining the expressions (\ref{lambdamu}) for the Lovelock couplings along the MDT,  with the stability constraints found in (\ref{stabwedge1})-(\ref{stabwedge2}). For each constraint we get a quartic equation on $\Lambda$ and we can study the number of real roots by examining the corresponding discriminants. For the lower boundary of the stability wedge, corresponding to the tensor channel, we get 
\begin{eqnarray}
\Delta_{s2}^{(h)}&=&-16 (d-1)^6 (K-1)^4 K^2 (d-3 K-1) (d-2 K-1)\times\nonumber\\ [0.7em]
&& \left[ (d (162 d-1529)+3768) K^4-(d (3 d (45 d-391)+1796)+4044) K^3\right.\\ [0.7em]
& & +(d (d (3 d (9d-56)-530)+3576)-504) K^2\nonumber\\ [0.7em]
&& \left. -4 (d-1) (d (d (2 d-21)-16)+300) K+8 (d-1)^2 ((d-2) d-12) \right] ~, \nonumber
\end{eqnarray}
where $s$ stands for stability and the number to the corresponding helicity of the channel considered. The $h$ indicates that we are just considering stability at the horizon, for the moment. 

For general values of $K$ this discriminant vanishes for $d=2K+1$ and $d=3K+1$. For $d<3K+1$ there is an interval of values for $\Lambda$ around $\Lambda=-K$ for which the tensor potential becomes unstable. This is simply due to the fact that the end point of the MDT is outside the stability wedge for $d<3K+1$. Then, by considering high enough $d$ for fixed $K$, the tensor channel becomes stable along our curve, close to the MDP.

We can proceed in a similar manner with the upper boundary of the naive stability region, given by the sound channel. The discriminant in this case is 
\begin{eqnarray}
\Delta_{s0}^{(h)}&=&-432 (d-2) (d-1)^6 (d-2 K-1) (d-K-1) (d-K) (K-1)^4 K^2 \times\\ [0.7em]
&&\left(((K-8) K+8) d^2-2 (K ((K-13) K+6)+8) d+K (20-K (23K+2))+8\right)\nonumber
\end{eqnarray}
and it yields zero whenever $d=2K+1$ and
\begin{equation}
d_{\pm}=\frac{K^3-13 K^2+6 K \pm \sqrt{(K-2)^2 K^2 \left(K^2+K-1\right)}+8}{K^2-8 K+8}~.
\label{stabd}
\end{equation}
The relevant root for the part of the curve we are interested in is actually $d_-$. For $d<d_-$ there is a unstable interval of $\Lambda$ values strictly inside $\Lambda\in(-\Lambda,-1)$ as can be seen in figure \ref{s0h_n3}. For $d=2K+1$ the MDP is at one of the edges of this interval. In the plot we can also check that for $d>d_-$ this unstable interval disappears leaving a perfectly stable trajectory until it reaches the MDP. The situation changes dramatically as we increase the order of the Lovelock theory though. This is true for $K<7$ -- for $K\geq7$ the denominator of $d_-$ changes sign and this critical value becomes negative and thus irrelevant. In other words, for $K\geq7$ the unstable interval remains for every dimension. These statements are illustrated by figures \ref{s0h_n3} and \ref{s0h_n8}.

\FIGURE{\includegraphics[width=0.6\textwidth]{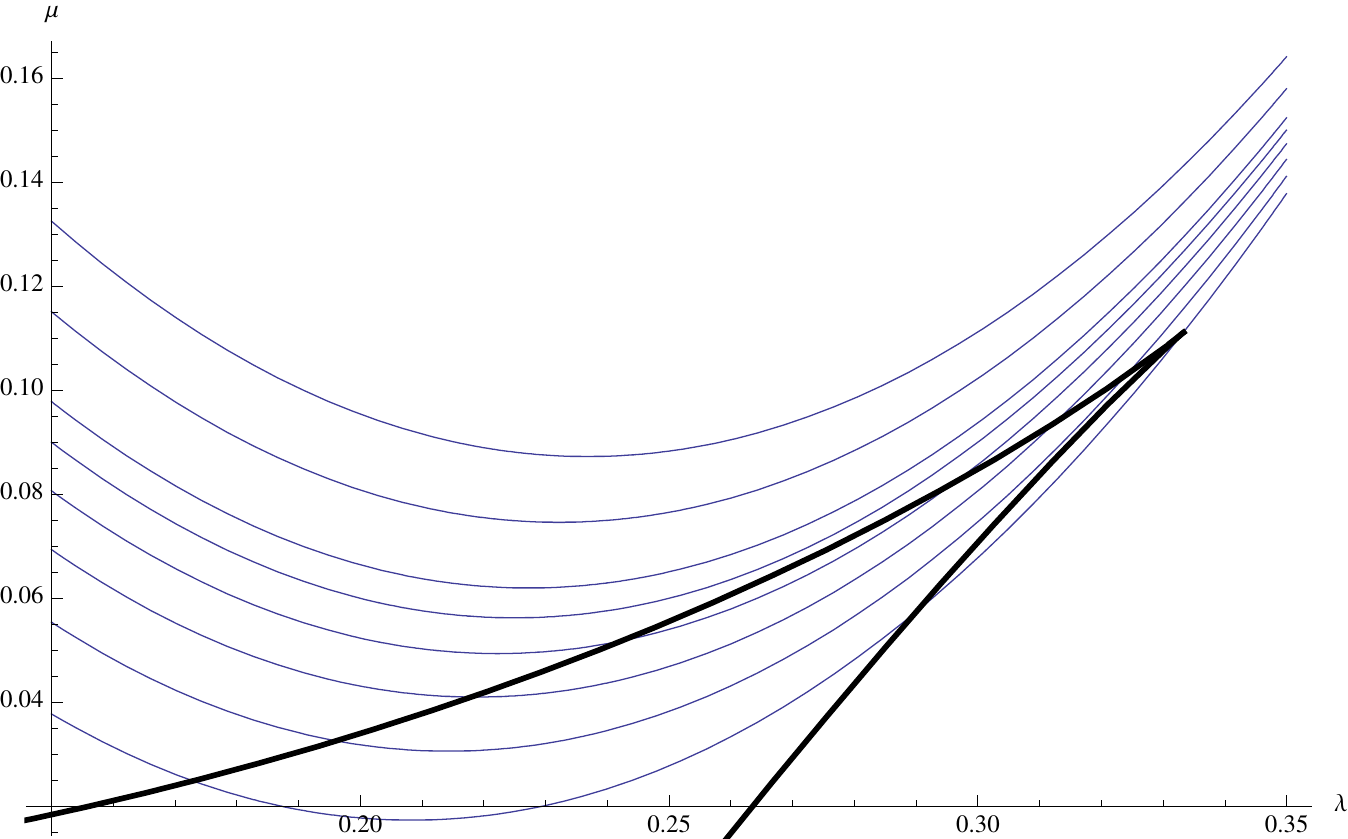}\caption{$K=3$ trajectory (in black) and $d=7,8,9,10,11,12,15,20$ helicity 0 stability constraint at the horizon (in blue). The unstable region is above the blue line. All the roots are in the figure in this case.}
{\label{s0h_n3}}}
\FIGURE{\includegraphics[width=0.6\textwidth]{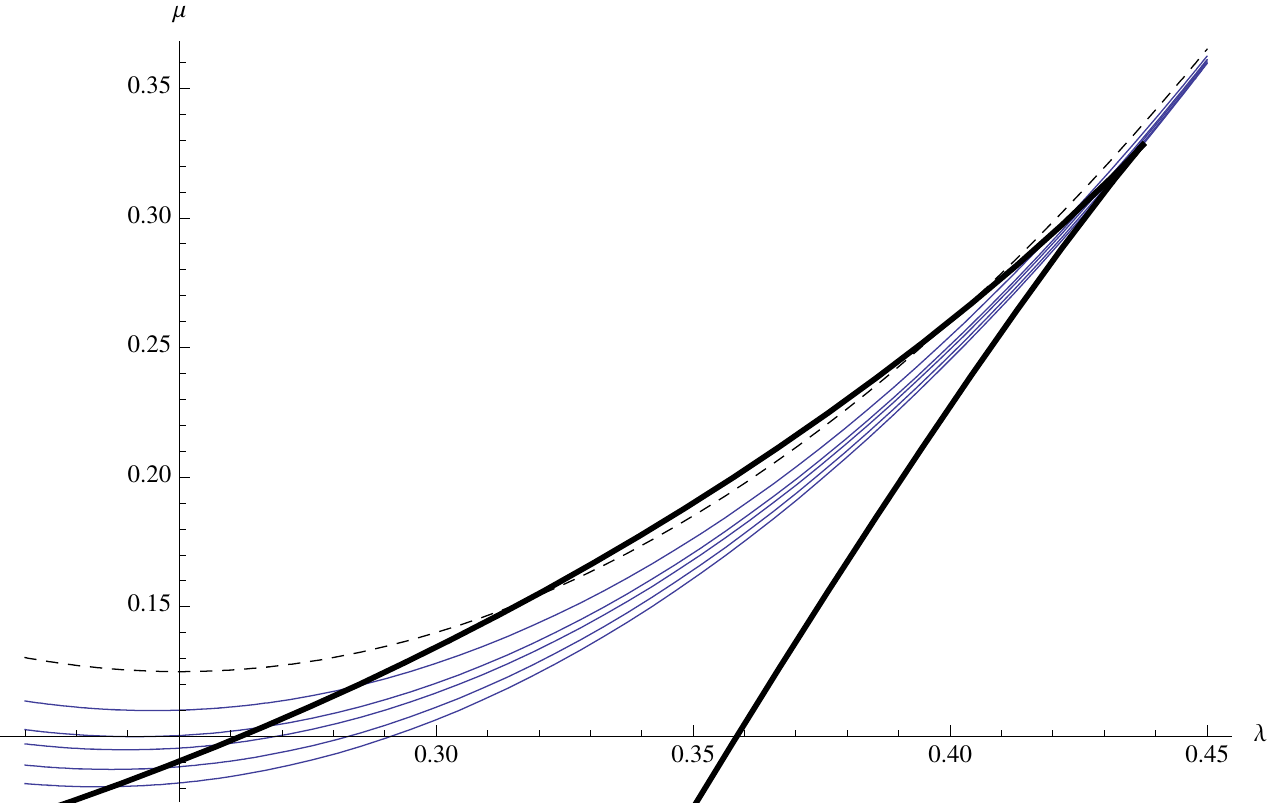}\caption{$K=8$ trajectory (in black) and $d=17, 20, 25, 30, 50,\infty $ helicity 0 stability constraint at the horizon (in blue and dashed black for $d\rightarrow\infty$). The unstable region is above the blue line. All the roots are in the figure in this case.}
{\label{s0h_n8}}}

\subsubsection{Full stability}

For $K>7$ it would appear that the sound channel is stable beyond the forbidden interval. This turns out not to be the case. As one goes beyond this interval the instability moves deeper in the bulk, so that there is always an instability in this channel. The full stability analysis is performed as follows. First notice that the two functions involved in the effective potentials, $F'(x)$ and $F''(x)$, have particularly simple expressions along the MDT:
\begin{eqnarray}
F'(x(g)) & = & \frac{(K-1) (g-\Lambda ) (\Lambda +1) \left(2 \Lambda ^2+K \left(-\Lambda
   ^2+g \Lambda +\Lambda +g\right)\right)}{\left(\Lambda ^2+K
   \left(-\Lambda ^2+g \Lambda +g\right)\right)^2}~,\\[0.9em]
F''(x(g)) & = & \frac{2 (K-1) (g-\Lambda ) \Lambda ^2 (\Lambda +1) (K+\Lambda )^2 ((K-1)
   \Lambda +g (\Lambda +1))}{\left(\Lambda ^2+K \left(-\Lambda ^2+g
   \Lambda +g\right)\right)^4}~.
\end{eqnarray}
We have removed the explicit dependence on the derivatives of the function $g(r)$ by using the polynomial equation (\ref{eqg}). For instance one gets:
\be
g'(r)=-\frac{d-1}{r}\frac{\Upsilon[g]}{\Upsilon'[g]} ~.
\ee
In this way we can analyze the potentials just knowing the range of values $g$ can take and avoiding solving the polynomial equation explicitly. 
In order to complete the stability analysis, now in the bulk of the geometry, we have to check if there is any value of $g\in[\Lambda,0]$ for which any of the potentials takes negative values. The appearance of any instability will be signaled with a change of sign of the discriminant of the corresponding polynomial equation $c_i^2=0$\footnote{This equation is not really polynomial but rational. However the denominator of the expression doesn't play any relevant role in this discussion and so we will restrict ourselves to the polynomial equation given by the numerator}. This is where a minimum of the potential touches the zero axis yielding a new double root of the equation. It is then enough to look for zeros of the discriminant and check if there is instabilities in the different disconnected regions. 

The full discriminant in the helicity zero case can be nicely written in terms of the corresponding discriminant of the horizon condition,
\begin{equation}
\Delta_{s0}=\frac{\Lambda ^{12} (K+\Lambda )^{12}}{(d-1)^{12} (\Lambda +1)^{12}}\Delta_{s0}^{(h)} ~,
\end{equation}
and then the appearance of instabilities in the bulk is controlled by the same discriminant analyzed in the precedent section. The $\Lambda$ dependent factor vanishes just at the maximally symmetric point as $\Lambda=0$ is never possible in the present case. From the previous analysis we know that there are two naively stable regions for $d<d_-$ and just a full naively stable trajectory as we go to higher dimensions. We have to consider separately the allowed values for $\Lambda$ on both sides of the unstable interval. 

For $K<7$ and any value of $\Lambda$ in between the forbidden interval and the MDP there is an interval of values for $g$ where the potential goes below zero. This instability disappears for $d>d_{-}$, at the same dimension as the naive unstable interval does (see left figure \ref{fullstab0_lowd}). Notice there's no instability either on the other allowed region as can be seen on the right hand side of figure \ref{fullstab0_lowd}.
\FIGURE{\includegraphics[width=0.45\textwidth]{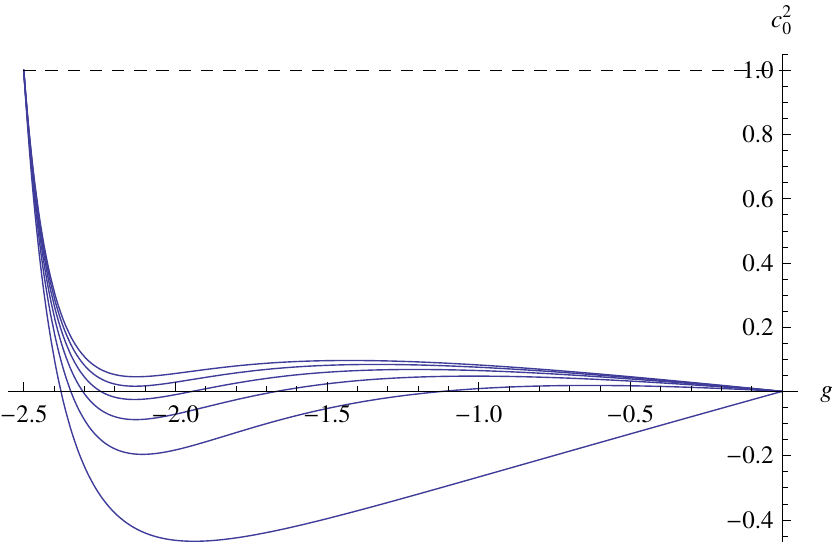}~~\includegraphics[width=0.45\textwidth]{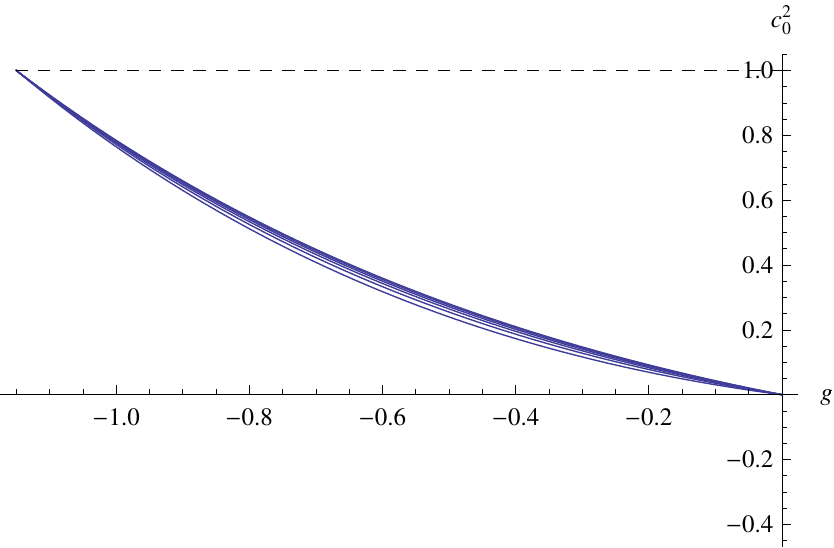}
\caption{Helicity zero potential for $K=3$ and $d=7,8,9,10,11,12$ to the left and to the right of the forbidden interval on $\Lambda$, $\Lambda=-2.5$ and $\Lambda=-1.15$ respectively.}
{\label{fullstab0_lowd}}}
For higher values of $K$ the instability in the allowed interval for $\Lambda$ close to the MDP never disappears. No matter how high the dimensionality some values of $g$ lead to negative values of the sound potential (see figure \ref{fullstab0_highd}). As one enters the stability wedge the instability simply moves away from the horizon and out towards the boundary of the geometry. 
\FIGURE{\includegraphics[width=0.45\textwidth]{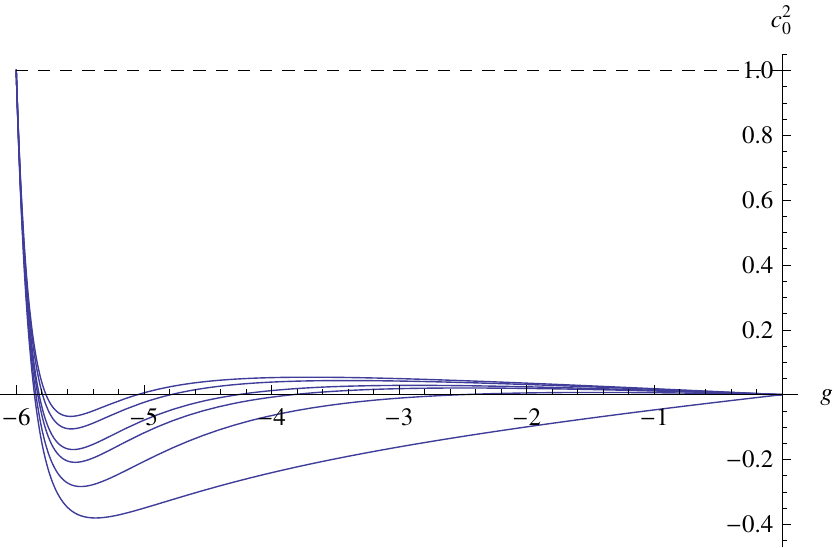}~~\includegraphics[width=0.45\textwidth]{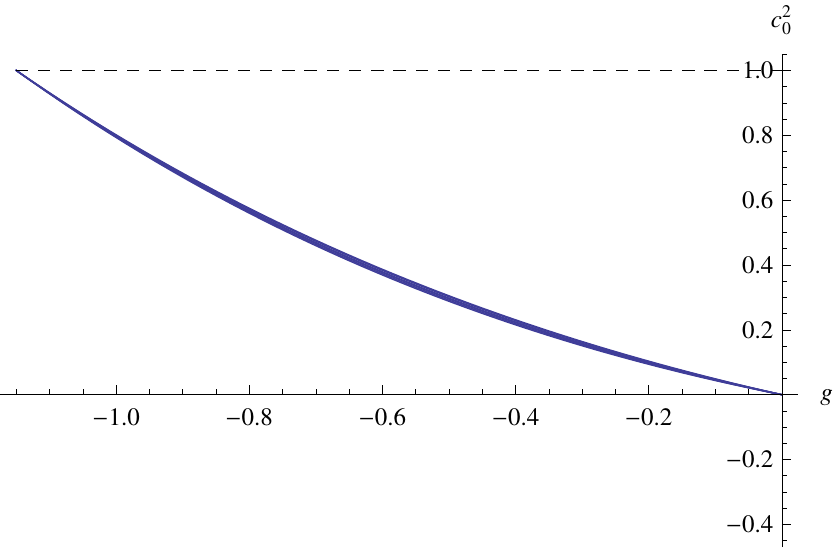}
\caption{Helicity zero potential for $K=8$ and $d=17,20,25,30,50,100$ to the left and to the right of the forbidden interval on $\Lambda$, $\Lambda=-2.5$ and $\Lambda=-1.15$ respectively.}
{\label{fullstab0_highd}}}

The same analysis can be done in the helicity two channel. There the full discriminant turns out to be
\begin{equation}
\Delta_{s2}=\frac{\Lambda ^{12} (K+\Lambda )^{12}}{(d-1)^{12} (\Lambda +1)^{12}}\Delta_{s2}^{(h)}~.
\end{equation}
In this case we just have one naively stable region for $\Lambda\in(-K,-1)$ and the tensor potential has been found to be stable everywhere there. Then the tensor channel is free of instabilities as long as it is so at the horizon (see figure \ref{fullstab2}).
\FIGURE{\includegraphics[width=0.6\textwidth]{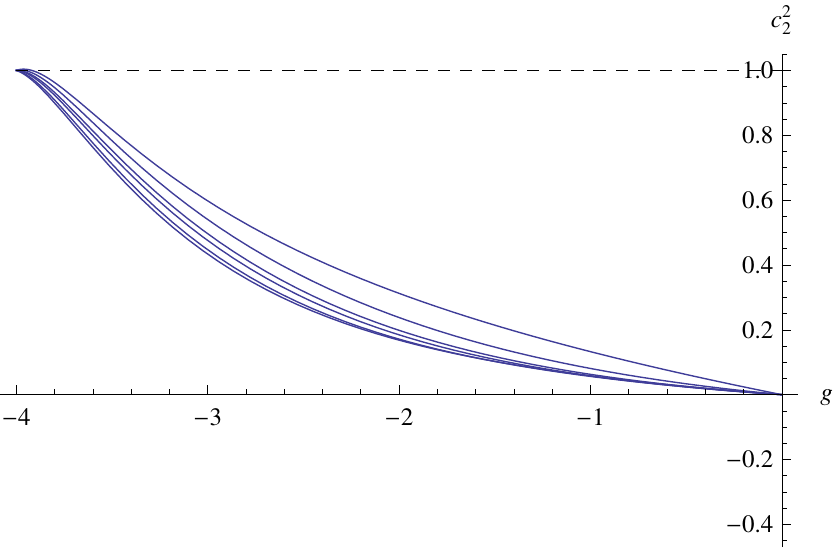}
\caption{Helicity two potential for $K=8$ and $d=17,20,25,30,50,100$ and cosmological constant $\Lambda=-4$. We find causality violation, but no instabilities.}
{\label{fullstab2}}}

Even though we haven't made any reference to it yet, we didn't forget the shear potential. The reason for this disregard is that the corresponding stability constraint is always less constraining than the helicity zero and two ones. In fact, it amounts just to $\lambda<\lambda_c$, or equivalently $\eta/s>0$, when the other two constraints are respected (see eq. \ref{potentialsF}).

All the features described in this section for the maximally degenerated trajectory can be easily observed in the cubic Lovelock case as we increase the dimensionality (see figures \ref{LL3}).  
\FIGURE{\includegraphics[width=0.5\textwidth]{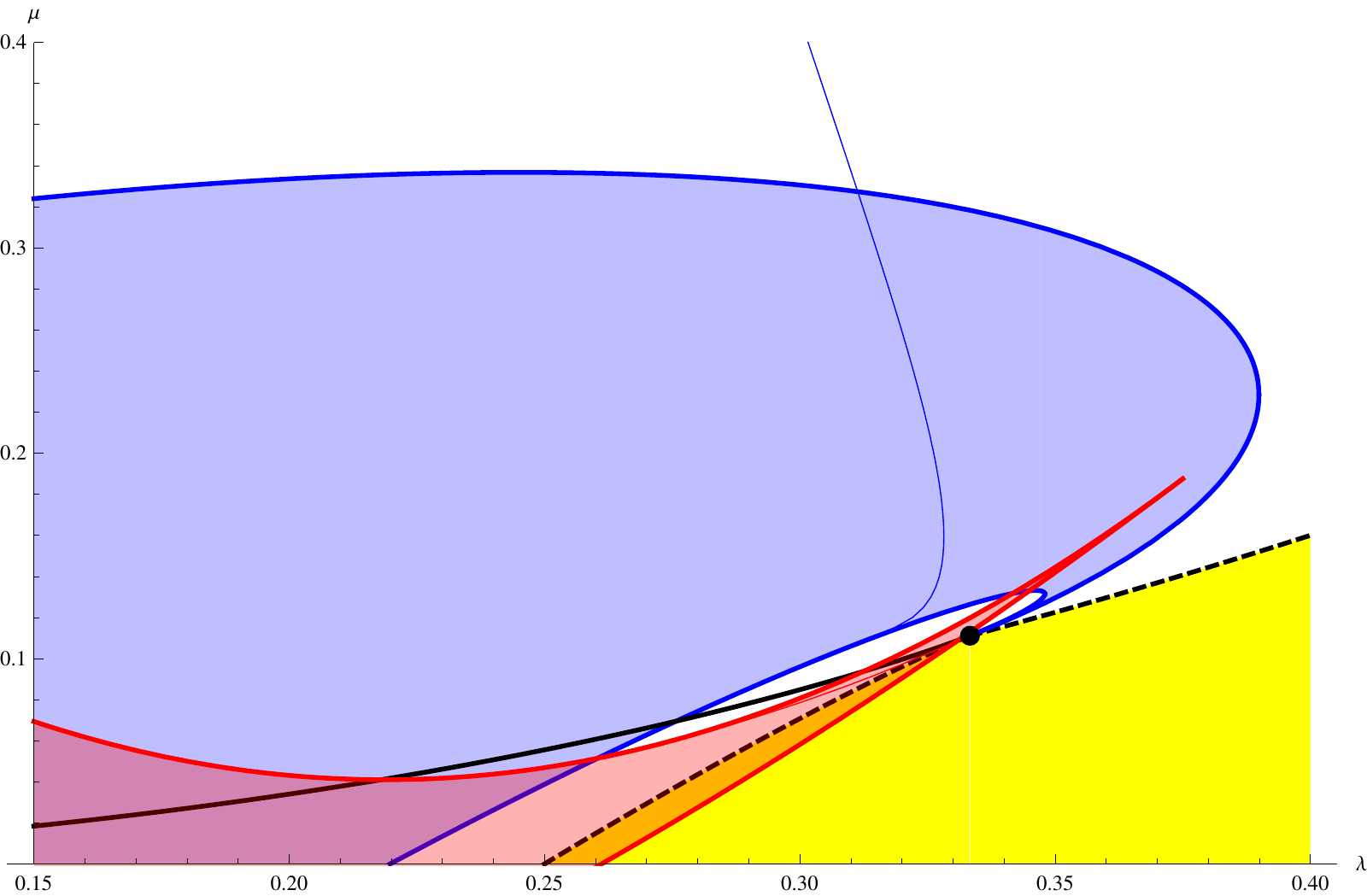}~~\includegraphics[width=0.4\textwidth]{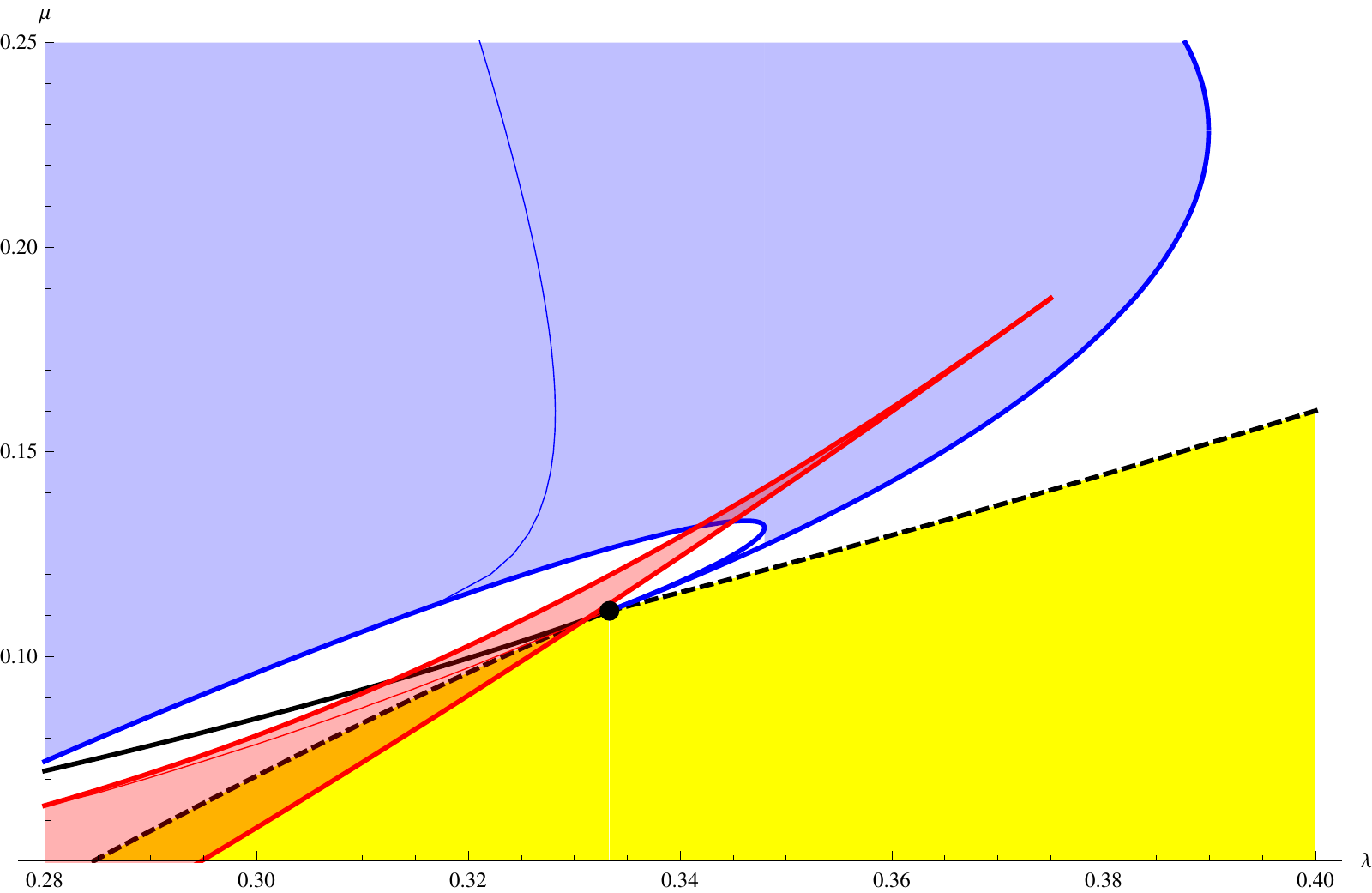}\\[0.5em]
\includegraphics[width=0.5\textwidth]{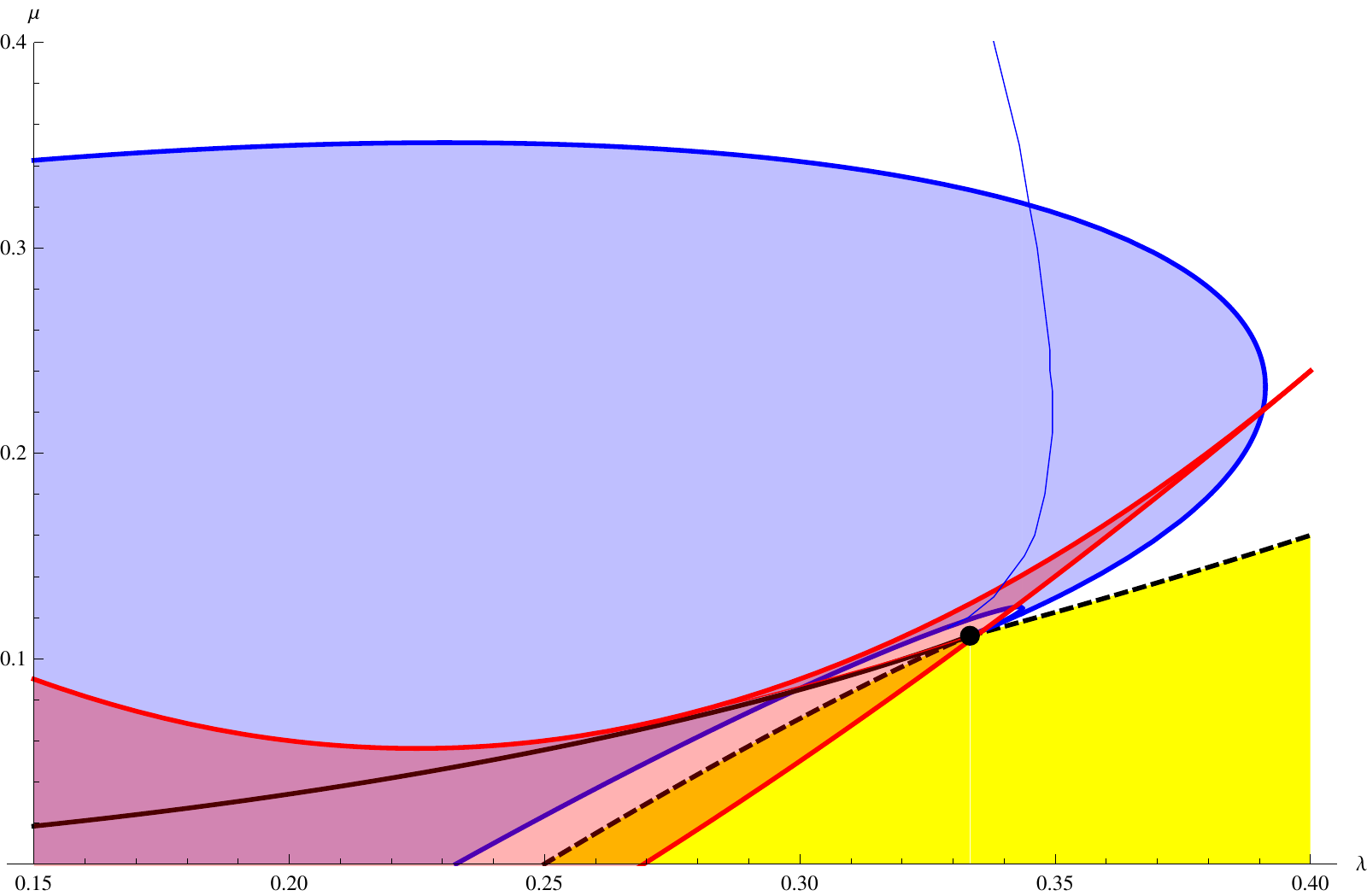}~~\includegraphics[width=0.4\textwidth]{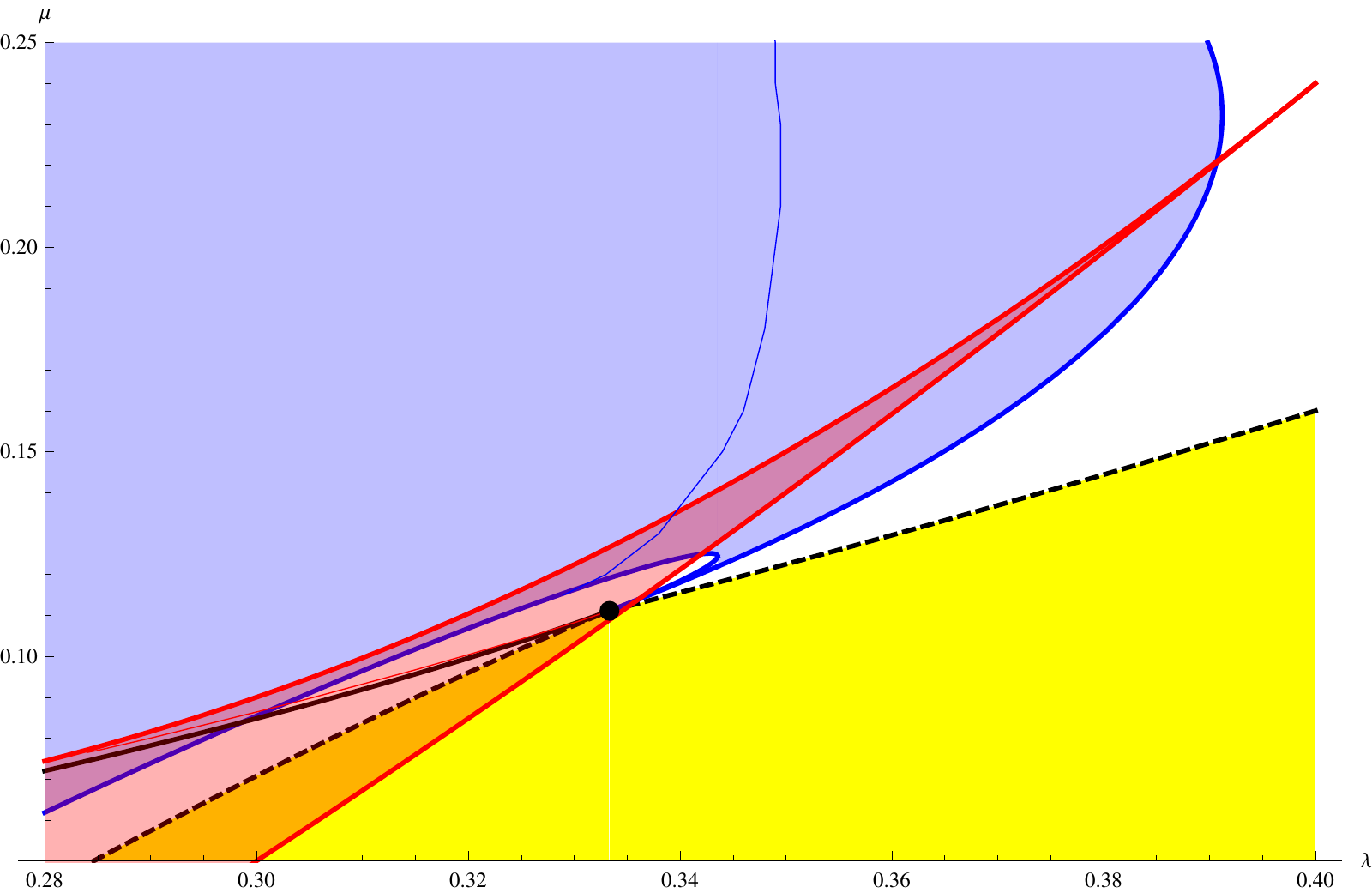}\\[0.5em]
\includegraphics[width=0.5\textwidth]{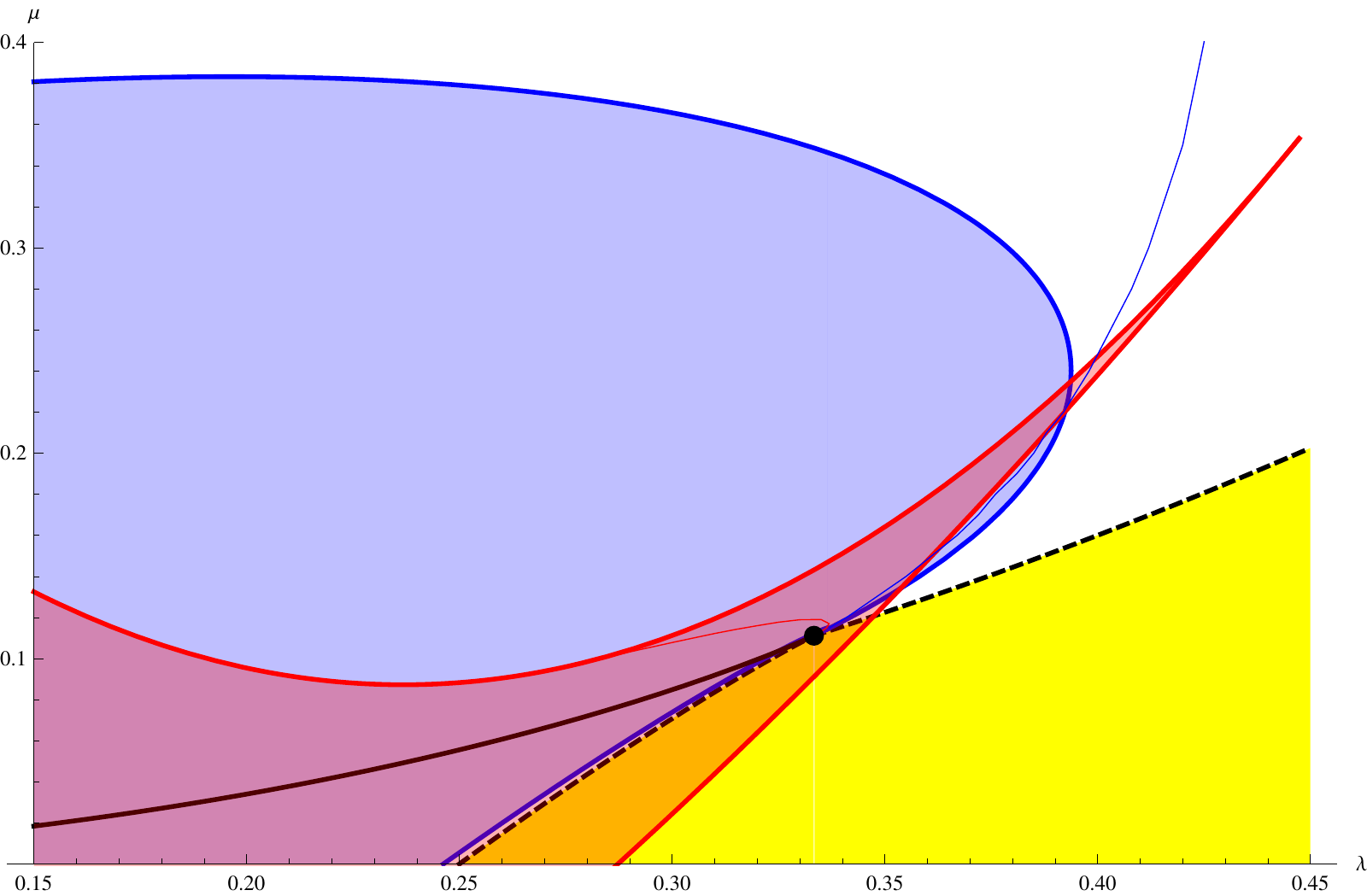}~~\includegraphics[width=0.4\textwidth]{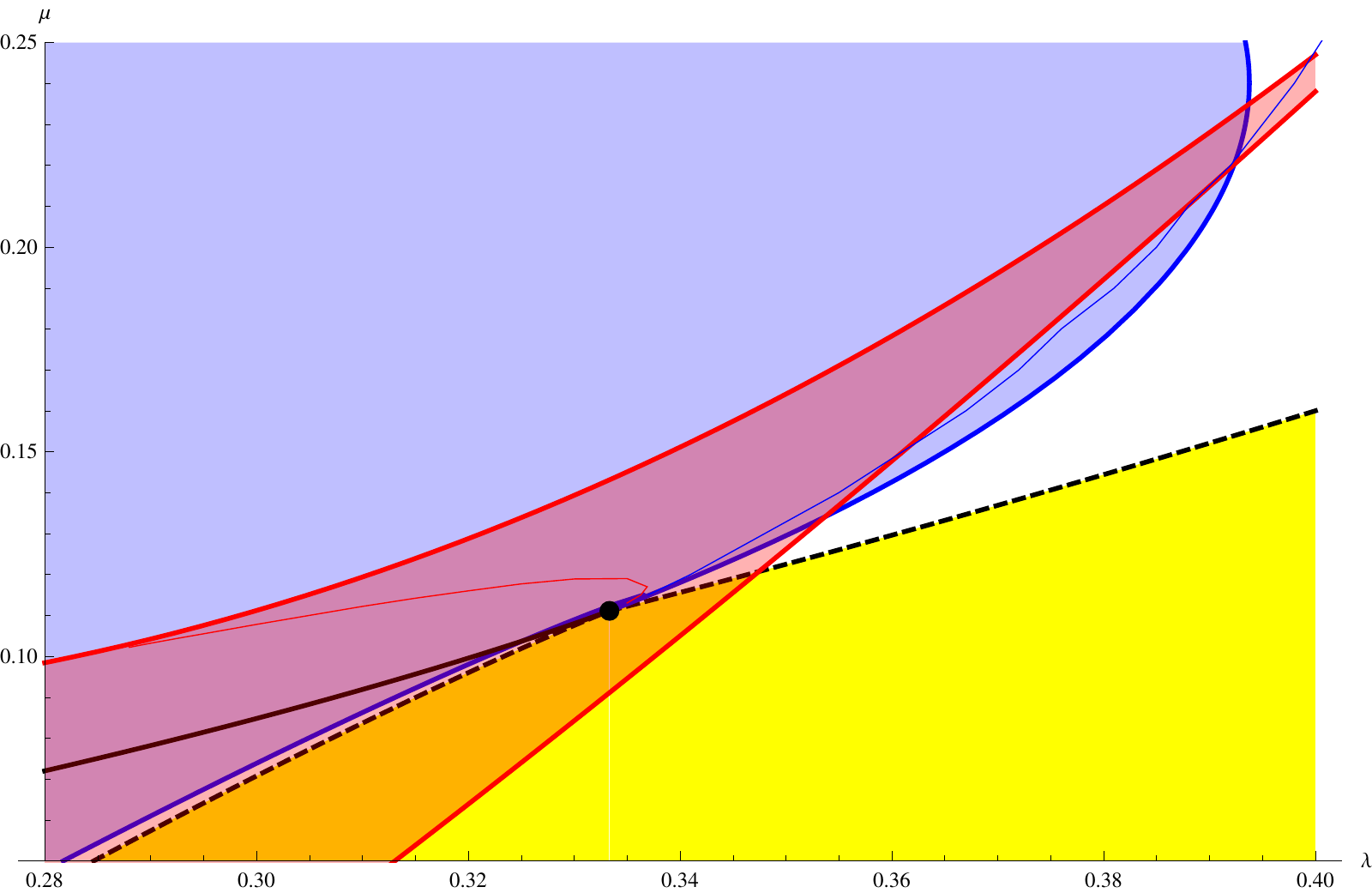}\\
\caption{All the different regions in the third order Lovelock case for $d=9,11,20$ (and zoom), from top to bottom. In black, the maximally degenerated line considered in the `full' causality and stability analysis. For $d<11$ the region of this curve closest to the maximally symmetric point (black dot) is unstable whereas for $d\geq11$ it becomes stable.}
{\label{LL3}}}

\subsection{Causality analysis}

Let's now try doing the same kind of analysis with the causality constraints. These are of the form
\begin{equation}
\Upsilon'[\Lambda]+\tau \Lambda\Upsilon''[\Lambda]=0
\end{equation}
For the MDT trajectory, solving this equation yields
\begin{equation}
\Lambda_{\star}=-\frac{K+2 (K-1) \tau }{1+2 (K-1) \tau}.
\end{equation}
Depending on the channel $\tau$ takes different values namely (see (\ref{genconstraints})),
\begin{eqnarray}
\tau_2&=&\frac{2(d-1)}{(d-3)(d-4)}\nonumber\\
\tau_1&=&-\frac{d-1}{d-3}\\
\tau_0&=&-\frac{2(d-1)}{(d-3)}\nonumber
\end{eqnarray}
In order to respect $\Lambda_{\star}<0$ we must have $\tau>-\frac{1}{2(K-1)}$ or $\tau<-\frac{K}{2(K-1)}$.
Then, the crossing point of the MDT with the helicity two constraint is between the Einstein-Hilbert point and the MDP, and the crossing with the other two constraints is before the Einstein-Hilbert point. The helicity two crossing happens at a particular value of the Gauss-Bonnet coupling, namely 
\begin{equation}
\lambda=\frac{K-1}{2K}\frac{K(K+2 (K-1) \tau)}{(K+2 (K-1) \tau_2)^2} ~.
\end{equation}
For the helicity two constraint the coefficient $\tau$ goes to zero as $d$ increases, that is, the value of $\lambda$ at the crossing point (the maximal value of $\lambda$ regarding causality) is the value at the  maximally symmetric point with an increasing extra factor that goes to one as $d\rightarrow\infty$. Then the crossing point approaches the MDP as $d$ increases. This is explained by the fact that the helicity two causality reduces to $C_T\geq 0$ for $d\to \infty$, and hence our trajectory can only cross it at the MDP itself.

The full causality analysis is quite more complicated that the stability one and not very enlightening. Here we will simply cite some partial results in the most relevant case, the helicity two one. In that case the equation $c_2^2(r)=1$ can be reduced to a polynomial equation on $g$ with discriminant
\begin{equation}
\Delta_{c2}= \Lambda^{20} (1+\Lambda)^{12} (K+\Lambda)^{12} \left(\Lambda-\Lambda_\star\right) \tilde{\Delta}_{c2}.
\end{equation}
The factor $\tilde{\Delta}_{c2}$ is a complicated expression which has zeroes at points irrelevant for our analysis. In particular, when restricted to the part of the curve connecting the Einstein-Hilbert point with the maximally symmetric point ($\Lambda\in(-K,-1)$) the discriminant has just one relevant  root at $\Lambda=\Lambda_{\star}$, exactly the point determined by demanding causality at the boundary. The situation is analogous for the other two helicities. Then, causality in the MDT reduces to causality at the boundary. In particular, close to the MDP the only constraint comes from the helicity two channel, as has been discussed.

In the same way as for the stability analysis, all the previously referred features can be seen in the third order case (see figures \ref{LL3}). 

\providecommand{\href}[2]{#2}\begingroup\raggedright\endgroup
\end{document}